\documentclass[fleqn,usenatbib]{mnras}

\usepackage[T1]{fontenc}
\usepackage{ae,aecompl}

\usepackage{graphicx}	% Including figure files
\usepackage{amsmath}	% Advanced maths commands
\usepackage{amssymb}	% Extra maths symbols

\usepackage{newtxtext,newtxmath}
\title[]{The origin of bulges and discs in the CALIFA survey: I. Morphological evolution.}

\author[J. M\'endez-Abreu et al.]{
J. M\'endez-Abreu,$^{1,2,3,4}$\thanks{E-mail: jairomendezabreu@gmail.com} A. de Lorenzo-C\'aceres, $^{1,2,5,6}$S. F. S\'anchez$^{3}$ 
\\
$^{1}$Instituto de Astrof\'isica de Canarias, Calle V\'ia L\'actea s/n, E-38205 La Laguna, Tenerife, Spain\\
$^{2}$Departamento de Astrof\'isica, Universidad de La Laguna, E-38200 La Laguna, Tenerife, Spain\\
$^{3}$Departamento de F\'isica y del Cosmos, Campus de Fuentenueva, Edificio Mecenas, Universidad de Granada, E18071, Granada, Spain\\
$^{4}$Instituto Carlos I de F\'isica Te\'orica y Computacional, Facultad de Ciencias, E18071, Granada, Spain\\
$^{5}$Departamento de F\'isica de la Tierra y Astrof\'isica, Universidad Complutense de Madrid (UCM), Plaza Ciencias 1, Madrid, E-28040, Spain\\
$^{6}$Instituto de F\'isica de Part\'iculas y del Cosmos (IPARCOS), Plaza Ciencias 1, Madrid, E-28040, Spain\\
$^{3}$Instituto de Astronom\'ia, Universidad Nacional Aut\'onoma de M\'exico, A.P. 70-264, 04510 M\'exico, D.F., M\'exico.\\
}

\date{Accepted XXX. Received YYY; in original form ZZZ}

\pubyear{2015}

\begin{document}
\label{firstpage}
\pagerange{\pageref{firstpage}--\pageref{lastpage}}
\maketitle

\begin{abstract}
This series of papers aims at understanding the formation and evolution of non-barred disc galaxies. We use the new spectro-photometric decomposition code, {\sc c2d}, to separate the spectral information of bulges and discs of a statistically representative sample of galaxies from the CALIFA survey. Then, we study their stellar population properties analising the structure-independent datacubes with the {\sc Pipe3D} algorithm. We find a correlation between the bulge-to-total ($B/T$) luminosity (and mass) ratio and galaxy stellar mass. The $B/T$ mass ratio has only a mild evolution with redshift, but the bulge-to-disc ($B/D$) mass ratio shows a clear increase of the disc component since redshift $z < 1$ for massive galaxies. The mass-size relation for both bulges and discs describes an upturn at high galaxy stellar masses ($\log{(M_{\star}/M_{\sun})} > 10.5$). The relation holds for bulges but not for discs when using their individual stellar masses. We find a negligible evolution of the mass-size relation for both the most massive ($\log{(M_{\star \rm ,b,d}/M_{\sun})} > 10$) bulges and discs. For lower masses, discs show a larger variation than bulges. We also find a correlation between the S\'ersic index of bulges and both galaxy and bulge stellar mass, which does not hold for the disc mass.
Our results support an inside-out formation of nearby non-barred galaxies, and they suggest that i) bulges formed early-on and ii) they have not evolved much through cosmic time. However, we find that the early properties of bulges drive the future evolution of the galaxy as a whole, and particularly the properties of the discs that eventually form around them.
\end{abstract}

\begin{keywords}
galaxies: bulge - galaxies: evolution - galaxies: formation - galaxies: structure - galaxies: disc
\end{keywords}

%%%%%%%%%%%%%%%%%%%%%%%%%%%%%%%%%%%%%%%%%%%%%%%%%%

%%%%%%%%%%%%%%%%% BODY OF PAPER %%%%%%%%%%%%%%%%%%
\section{Introduction}

Galaxies are complex systems with an intricate combination of different structural components such as bulges, discs, and bars. The relative contribution of these structures to the galaxy luminosity, or mass, define our morphological classifications, which have been the subject of numerous studies along the last century \citep{hubble36,devaucouleurs64,buta15}. Many physical properties of galaxies, such as gas content, stellar age, and star formation rate (SFR) are known to correlate with morphology. Therefore, studying the cosmic evolution of the main structures of galaxies is key to understand the physical mechanisms driving the morphological evolution of galaxies \citep{clauwens18, tacchella19}.

In the simplest scenario, disc galaxies are made up of a central bulge and an outer disc. Galactic discs are thought to form within dark matter haloes with high angular momentum, and a quiet recent assembly history, as a consequence of angular momentum conservation during the dissipational collapse of gas \citep{fallefstathiou80,mo98}. On the other hand, bulges are generally seen as the slowly-rotating result from merger events \citep{cole00}. However, more recent, sophisticated simulations are challenging this simple view \citep[e.g.,][]{sales12}.

An early origin of galactic discs, driven by the infall of gas in a rotating dark matter halo, has been the preferred scenario for the formation of this component during decades \citep{fallefstathiou80}. In this framework, the central regions of discs reach the necessary gas surface mass density to form stars earlier in time than the outer parts, naturally resulting in an inside-out mass growth \citep{brook06}. However, this idealised picture is more complex in a hierarchical model of galaxy formation where i) galaxy mergers can destroy (or thicken) the initial discs \citep{steinmetznavarro02}
and ii) gas does not keep all its initial angular momentum, thus producing discs which are too small compared to those observed in nearby galaxies \citep{navarrowhite94,sommerlarsen99,donghiaburkert04}. Therefore, at least for a fraction of disc galaxies, a later formation must be invoked. From the structure formation point of view, alternative dark matter properties (such as Warm Dark Matter) will induce that structure formation occurs later \citep{sommerlarsendolgov01}. More related with the physics of baryons, the effect of stellar or active galactic nuclei (AGN) feedback might also prevent the gas from cooling until relatively late times $z<1$ \citep{weil98,thackercouchman01}. More recently, hydrodynamical simulations have demonstrated that after a major merger of gas-rich galaxies, a new disc can be formed out of the remaining gas not consumed in the initial starburst \citep{hopkins09}. In these delayed formation scenarios, the relation between the initial angular momentum of the halo and that of the disc is expected to be erased. 

A variety of pathways for bulge formation has been proposed in the literature. At high redshift, when the gas mass fraction in galaxies was higher than at present days, the formation of galaxy bulges was mainly driven by highly dissipative processes. Some of the proposed scenarios are similar for bulges and ellipticals, such as the direct monolithic collapse of protogalactic gas clouds \citep{eggen62,larson74} or the major mergers of gas-rich galaxies \citep{kauffmann96,hopkins09,zavala12,avilareese14}. Both of these mechanisms imply (in order to not end up in an elliptical galaxy) that the stellar disc is formed after the bulge is already in place. This would be in agreement with the idea of an inside-out mass growth for disc galaxies \citep{aumerwhite13,gonzalezdelgado15}, but some works have questioned that the frequency of major mergers might not be enough to be the primary formation mechanism of bulges and ellipticals \citep{kitzbichler08}. Other dissipative mechanisms have been proposed where the disc might form before, or at least concomitantly to, the bulge. At high redshift, primordial gas-rich galaxy discs are highly turbulent with star formation occurring in massive clumps \citep{abraham96,elmegreen04,hinojosagoni16}. The coalescence of these giant clumps as they move to the galaxy center due to dynamical friction has been demonstrated to create new bulges \citep{noguchi99,immeli04,bournaud07,ceverino15}. However, the role of clumps in bulge formation is still not clear since their survival strongly depends on the feedback implementation \citep{mandelker17,oklopcic17}. Another bulge formation scenario at high redshift is related to a rapid gas inflow injected to the galaxy center directly from the surrounding halo \citep{scannapieco09,zolotov15,tacchella16}. In this scenario, spheroids tend to form when the spin of newly-accreted gas is misaligned with that of the host galaxy, leading to episodic formation of stars with different kinematics that cancel out the net rotation of the galaxy \citep{sales12}. At lower redshifts, galaxy bulges might continue to grow from: i) stars already present in their hosting discs through radial migration \citep{minchev10} or dissolution of bars \citep{guo20}, ii) ex-situ stars accreted from satellites during minor merger events \citep{aguerri01, elichemoral06,guedes13,rodriguezgomez17}, or iii) in-situ new stars created from the inflow of gas from the outer disc to the galaxy center due to the gravitational torque exerted by stellar bars \citep{kormendykennicutt04,athanassoula05}. All these latter pathways for bulge formation require longer timescales to modify/create bulges and they are generally referred to as secular processes. 

The diversity of formation and evolution scenarios for bulges has been generally condensed into two broad observational classes with different characteristics: classical and disc-like bulges \citep{kormendykennicutt04, athanassoula05}. Classical bulges are those following surface-brightness distributions with a S\'ersic index $n > 2$ and bulge-to-total ($B/T$) luminosity ratio $B/T > 0.2$, they appear rounder than their surrounding discs \citep{mendezabreu10}, and their stellar kinematics is dominated by random motions that generally satisfy the fundamental plane (FP) correlation \citep{bender92,falconbarroso02,aguerri05}. The stellar populations of classical bulges show similarities with those of ellipticals of the same mass. In general, they are old and metal-rich with a short formation timescale \citep[see][]{sanchezblazquez16}. On the other hand, disc-like bulges are oblate ellipsoids \citep{costantin17a} with apparent flattening similar to their outer discs, surface-brightness distributions well fitted with a S\'ersic profile of index $n < 2$ and $B/T < 0.35$ \citep{fisherdrory08}. Their kinematics is dominated by rotation in diagrams such as the $v/\sigma$ vs. $\epsilon$ \citep{kormendykennicutt04} and they are identified as low-$\sigma$ outliers of the Faber-Jackson relation \citep{faberjackson76}. Disc-like bulges are also usually dominated by young stars, with the presence of gas and possible recent star formation \citep{fisherdrory16}. Despite this apparently clear separation in their observed properties, a number of studies have demonstrated that: i) several of the previous observables do not show a clear dichotomy, making it difficult to establish the limits between both bulge types and ii) using different diagnostics generally lead to different classifications producing heavily contaminated samples \citep{costantin18,mendezabreu18,costantin20}. This situation might be caused (or enhanced) by the discovery that bulges in at least some disc galaxies are indeed composite systems, i.e., a single galaxy can host a (kinematically hot, spheroidal) classical bulge and a (kinematically cool, flattened) disc-like bulge \citep{mendezabreu14,erwin15}. All previous caveats on bulge classification, together with the fact that the large variety of bulge formation paths envisioned in simulations are only coarsely captured with the current observational division, have prompted us to avoid such a separation in this study.

In the last decade, the study of galaxy bulges in particular, and disc galaxies in general, has strongly benefited from: i) developments on photometric techniques to isolate the different stellar components of galaxies, which have become more sophisticated ({\sc gasp2d}, \citealt{mendezabreu08a}; {\sc galfit}, \citealt{peng10}; {\sc imfit}, \citealt{erwin15b}). The generalisation of the idea that only detailed multi-component photometric decompositions are suitable to understand the origin of bulges has contributed to these improvements in recent years \citep{gadotti09,mendezabreu17,delorenzocaceres19b}; ii) the advent of integral field spectroscopy (IFS), that has given access to the spatially resolved properties of individual components, therefore allowing for a deeper understanding of their kinematic and stellar population properties. However, 
a general problem hindering our advance on understanding the formation and evolution of bulges and discs galaxies relies on the necessity of new techniques to analyse the data. In particular, there is a strong need for new ways of separating, spectroscopically, the stellar structures shaping the galaxies avoiding issues with contamination due to overlapping. Recently, new algorithms have explored different approaches to work out this problem. We classify them here in three classes: i) spectro-photometric decompositions. In this technique IFS data is understood as a series of two-dimensional images that be decomposed into their structural components using standard photometric decompositions codes \citep{johnston14,johnston17,johnston21,mendezabreu19,mendezabreu19b,barsanti21}, ii) kinematic decompositions. The structural components are assumed to have different kinematic distributions that are directly fitted, or derived, from the spectra \citep{coccato11,tabor17,coccato18,mehrgan19}; iii) Schwarzchild dynamical modelling. The galaxy stellar kinematics is modelled using a variety of stellar orbits. These are then analysed, generally in terms of their angular momentum, to identify structures within the galaxy \citep{zhu18,zhu18b,poci19}. These new methods are providing more accurate constraints to numerical simulation models of disc galaxy formation.

This paper is the first of a series devoted to analyse, with unprecedented detail, a statistically representative sample of bulges and discs in the nearby Universe. To this aim, we have applied our recently developed spectro-photometric decomposition code \citep[{\sc c2d}; ][]{mendezabreu19} to a sample of photometrically classified bulge-to-disc galaxies observed within the CALIFA IFS survey \citep{sanchez16a}. We have explicitly removed barred galaxies from our sample to {\it simplify} the interpretation of our results. Nonetheless, we appreciate they constitute a important channel for bulge formation, and have an strong impact on the evolution of galactic discs, so they will be thoroughly studied in a separate paper. As stated before, in this series of papers we refrain from a pre-defined separation of bulge types and study both bulges and discs spectro-photometric properties as a continuous population. In this first paper, we focus on understanding the morphological evolution of disc galaxies, to this aim we explore the relations between the main photometric properties of bulges and discs, as derived from typical photometric decompositions, their stellar masses obtained from spectral synthesis modelling and, whenever possible, their time evolution using the derived star formation histories (SFHs).

This paper is organised as follows. Sect~\ref{sec:sample} describes the sample of photometrically classified bulge-to-disc galaxies from the CALIFA survey. Sect~\ref{sec:c2d} highlights the main features of our new methodology to extract the spectro-photometric properties of bulges and discs using IFS data and its analysis to derive their stellar population properties. Sect~\ref{sec:results} shows the main relations between photometric properties and galaxy mass, as well as their time evolution whenever possible. Sect~\ref{sec:discussions} and Sect~\ref{sec:conclusions} summarise our main conclusions. Throughout the paper we assume a flat cosmology with $\Omega_m$ = 0.27, $\Omega_{\lambda}$ = 0.73, and a Hubble constant $H_0$ = 71 km s$^{-1}$ Mpc$^{-1}$.

%----------------------------------------------
%----------------------------------------------
\section{CALIFA sample of bulge and disc galaxies}
\label{sec:sample}

The sample of galaxies analysed in this work is drawn from the CALIFA data release 3 \citep[DR3;][]{sanchez16b}. This data release comprises 667 galaxies covering a wide range of stellar masses and Hubble types. \citet{mendezabreu17} carried out a multicomponent multiband ($g-$, $r-$ and $i-$band) photometric decomposition of 404 galaxies present in the CALIFA DR3 using SDSS imaging. This includes all galaxies but those with high inclination ($i > 70$ degrees) or in interaction \citep[see][for more details]{mendezabreu17}. From this study, we take the 70 galaxies that were successfully fitted using only a central bulge (represented with a S\'ersic profile) and an outer disc (modelled with a single exponential profile), as well as the 64 galaxies that were fitted using a central bulge and an outer broken exponential disc, i.e., discs profiles of types II (down-bending) and III (up-bending) following the definition of \citet{erwin05}. In order to avoid possible issues with the fit around the break radius, we discard those galaxies with $r_{\rm break} < 25$ arcsec, therefore remaining with 57 galaxies with a well-behaved bulge and disc inside the CALIFA field-of-view (FoV). We finally include in the sample the 32 early-type galaxies classified as {\it unknown} in \citet{mendezabreu18} and already analysed using {\sc c2d} in \citet{mendezabreu19}. This latter sample represents early-type galaxies for which our photometric approach is not able to classify them in either a simple S\'ersic or S\'ersic+Exponential fit, as both options return the same statistical solution. These galaxies are analysed here using their two-component (bulge and disc) best fit. We run {\sc c2d+Pipe3D} over this sample of 159 galaxies finding that for 30 of them the code does not converge, mainly due to their low $B/T$ ratio. Therefore, the final sample used in this paper comprises 129 non-barred disc galaxies with a photometric bulge: 58 bulge-to-disc, 40 bulge-to-disc with a break, and 31 early-type galaxies. In addition, we also use 41 photometrically classified elliptical galaxies for comparison \citep[see][for details]{mendezabreu19}.

The CALIFA survey \citep{sanchez16} observed all galaxies using the PMAS/PPaK instrument with a FoV of 74\arcsec $\times$ 64\arcsec. After a three pointing dithering pattern, the reconstructed datacubes cover up to $2.5 \times r_{\rm e,g}$ (galaxy effective radius) for 90\% of the sample \citep{walcher14}. The wavelength range and spectroscopic resolution for the adopted V500 setup (3745-7500 \AA, $R \sim 850$) are perfectly suited to study the properties of stellar populations and ionised-gas emission lines. The typical spatial resolution of the datacubes is full width at half-maximum (FWHM) $\sim$ 2.5 arcsec, corresponding to $\sim$ 1 kpc at the average redshift of the survey \citep{garciabenito15}.

%------------------------------------------------------
%------------------------------------------------------
\section{{\sc c2d}+{\sc Pipe3D} analysis of the sample}
\label{sec:c2d}

The spectro-photometric decomposition of the CALIFA galaxies into a bulge and disc is performed using {\sc c2d} \citep{mendezabreu19}. This new methodology allows us to separate the spectral contribution of each structural component providing an independent datacube for both bulge and disc. An extensive description of the code and its reliability is presented in \citet{mendezabreu19}. In brief, the application of {\sc c2d} to the CALIFA data is based on the idea that datacubes can be worked out as a sequence of quasi-monochromatic 2D images at different wavelengths. Thus, standard 2D photometric decomposition techniques are able to isolate the photometric contribution of both bulge and disc. In {\sc c2d}, the photometric decomposition engine is provided by {\sc gasp2d} \citep{mendezabreu08a,mendezabreu14}, a code that has been extensively tested on different galaxy samples with multiple structures \citep[e.g.,][]{delorenzocaceres19a,delorenzocaceres19b,delorenzocaceres20}. Moreover, {\sc gasp2d} was used in \citet{mendezabreu17} to perform a multiband photometric decomposition of the CALIFA galaxies using SDSS imaging. This is of particular importance in our case since the structural parameters of bulges and discs in {\sc c2d} are anchored to those derived using SDSS due to the coarse spatial resolution of the CALIFA datacubes. In fact, we note here that performing a completely free bulge-to-disc fitting directly on the CALIFA datacubes should be avoided \citep[see][ for the effect of spatial resolution on bulge parameters]{mendezabreu18}. Rearranging the best fitted intensity values for each quasi-monochromatic image into a datacube we are able to recover the characteristic spectrum for each component. In addition, {\sc c2d} provides an independent datacube (with spatial and spectral information) for each component. To do this, for each quasi-monochromatic image, the $B/T$ and the disc-to-total ($D/T=1-B/T$) ratios are computed spaxel-wise. Each fraction is then multiplied by the observed CALIFA datacube in that spaxel and wavelength producing independent bulge and disc datacubes. It is worth reminding that our spectro-photometric decomposition relies on the common assumptions that: i) bulges and discs can be well described using a S\'ersic and exponential analytical function, respectively (but see \citealt{breda20} for possible issues when dealing with late-type galaxies); and ii) galaxy structure changes smoothly with wavelenght. Further details on the specific application of {\sc c2d} to CALIFA data are presented in \citet{mendezabreu19}.

The second step of our analysis consists of deriving the stellar population and ionised gas properties of bulges and discs in our sample. We use the {\sc Pipe3D} pipeline \citep{sanchez16b}, which is specifically designed to extract the stellar population and ionised-gas properties from IFS data and it has been extensively tested on CALIFA data \citep[e.g.,][]{sanchez15,sanchez16a}. {\sc Pipe3D} adopts the GSD156 library of simple stellar populations from \citet{cidfernandes13}, that comprises 156 templates covering 39 stellar ages (from 1 Myr to 13 Gyr) and four metallicities (Z/Z$_{\sun}$ = 0.2 dex, 0.4 dex, 1 dex, and 1.5 dex). The best-fit stellar-population model spectra to the galaxy continuum is subtracted from the original cube to create a gas-pure cube including the ionised-gas emission lines only. Emmission lines are then measured using a set of Gaussian functions and subtracted to the original spectra to perform the analysis in an iterative sequence. The main data products obtained from {\sc Pipe3D} include luminosity/mass weighted ages and metallicities, star formation histories, and intensity maps of strong emission lines for both components. A Salpeter initial mass function (IMF) was adopted in the stellar population analysis \citep{salpeter55}. We refer the reader to the presentation paper of {\sc Pipe3D} \citep{sanchez16a} for a detailed description of its application to CALIFA.

\begin{figure*}
\begin{center}
\includegraphics[width=0.49\textwidth]{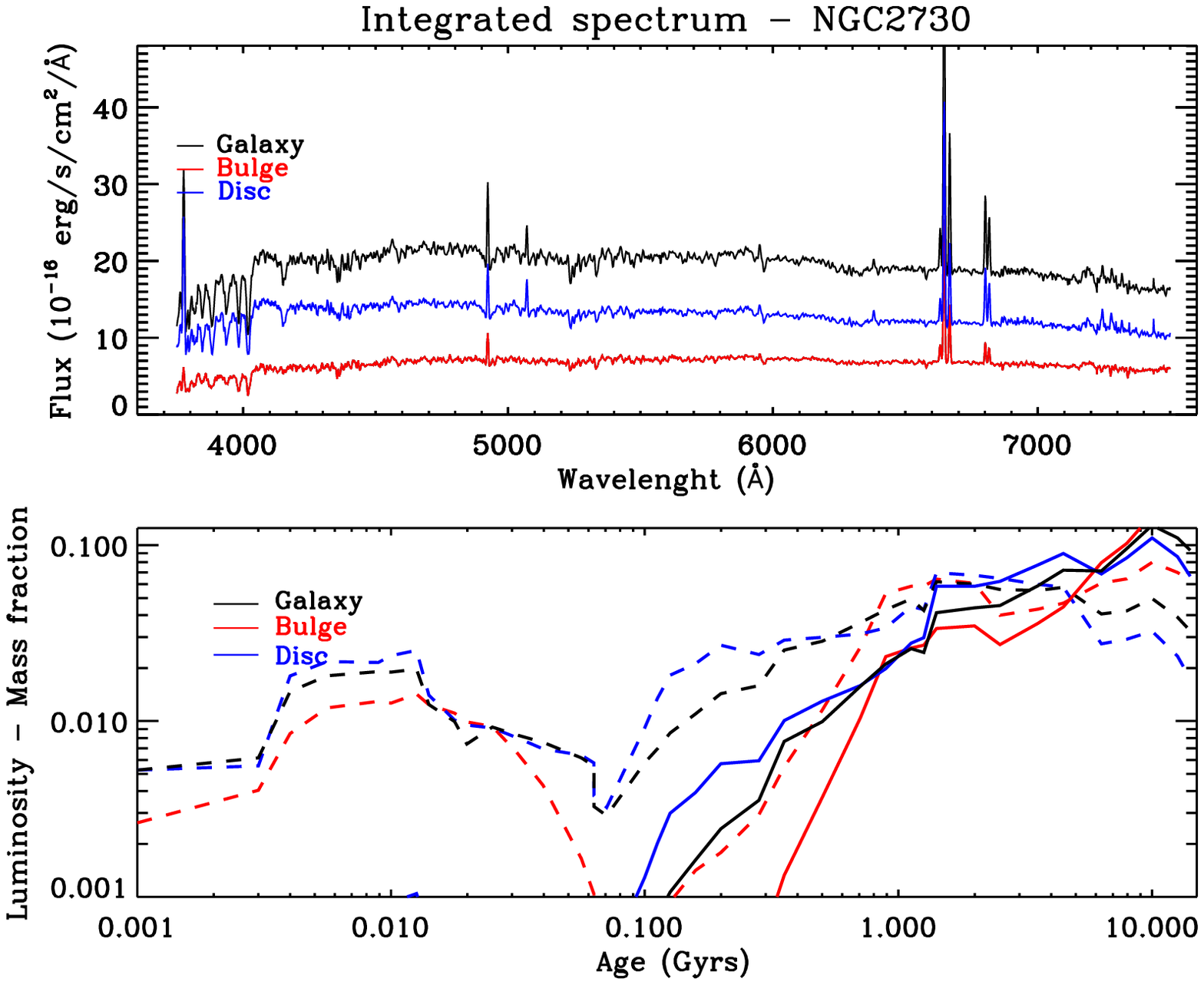}
\includegraphics[width=0.49\textwidth]{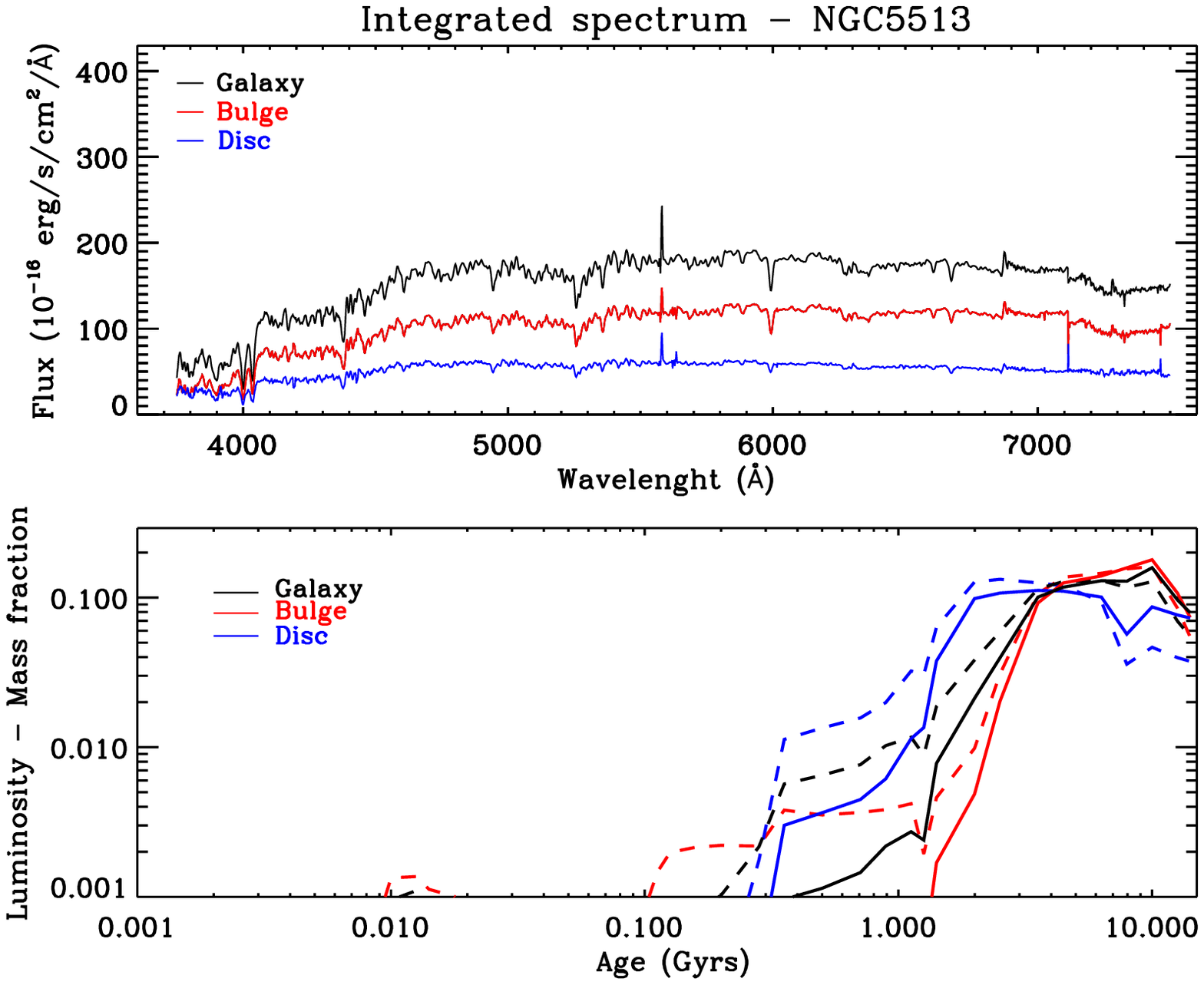} 
\caption{Left panels. NGC2730 ($\log{(M_{\star}/M_{\sun})}$ = 9.8  and $B/T$ = 0.17). Right panels. NGC 5513 ($\log{(M_{\star}/M_{\sun})} = 11.1$  and $B/T$ = 0.61). Upper panels. Integrated spectrum over one effective radius of the galaxy ($r_{\rm e, gal}$) for the whole galaxy (black), bulge (red), and disc (blue). Bottom panels. Luminosity (dashed lines) and mass (solid lines) fraction of stars contributing to a given stellar population age. Different colors as in the upper panels. Fractions are also computed within 1 $r_{\rm e, gal}$.}
\label{fig:integratedSFH}
\end{center}
\end{figure*}

Figure~\ref{fig:integratedSFH} shows the spectra (upper panels) and the fraction of light/mass (bottom panels) contributed by stars of different ages (marginalised over all possible metallicities) for the galaxy, bulge, and disc spectra of both NGC2730 and NGC5513. All these quantities are integrated within an ellipse with semi-major axis of one effective radius of the galaxy ($r_{\rm e, gal}$, see Table~\ref{tab:properties}) and oriented with the ellipticity and position angle of the outer disc. They are shown as an example of a typical low-mass and low-$B/T$ galaxy (NGC2730) and a high-mass and high-$B/T$ galaxy (NGC5513) in our sample.

%--------------------------------------------------------
%--------------------------------------------------------

%------------------------------------------------------
%------------------------------------------------------
\section{Results}
\label{sec:results}

In this section,  we first describe the general properties of our sample and define the stellar mass ranges where it can be considered statistically representative of the general population of local galaxies. We then focus on the main relations obtained between the photometric properties of our bulges ($r_{\rm e}$, $n$) and discs ($h$) with the stellar mass ($M_{\star}$) derived for both the whole galaxy and the individual components. We provide the cosmic evolution of these properties for our sample when possible.

%------------------------------------------------------
\subsection{Spectro-photometric properties of the sample}
\label{sec:generalprop}

\begin{figure*}
    \centering
    \includegraphics[bb=54 510 558 670,width=0.99\textwidth]{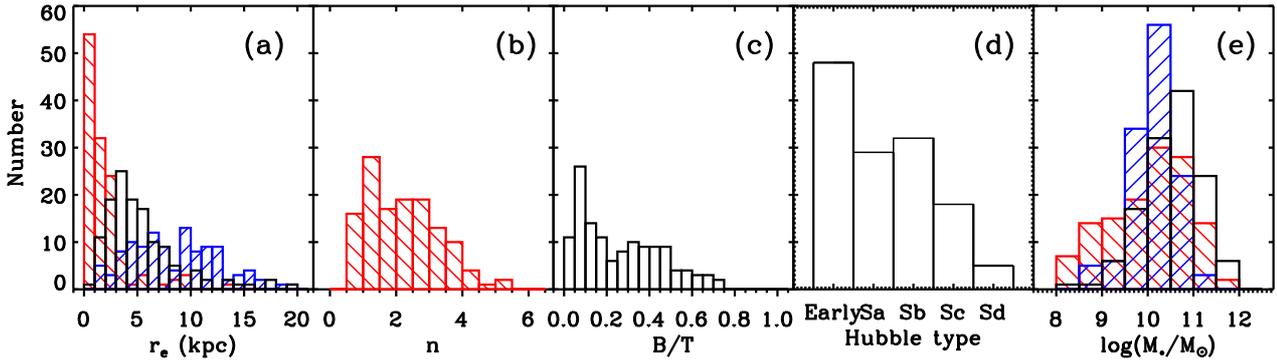}
    \caption{(a) Distribution of effective radii for bulges (red), discs (blue), and the whole galaxy (black). The values for bulges and discs are from the $r-$band photometric decompositions of \citet{mendezabreu17}. The effective radii of the galaxies are obtained from the $r-$ band analysis performed in \citet{walcher14}. (b)  Distribution of S\'ersic indices for our bulges and (c) bulge-to-total light ratio in the $r-$band from photometric decompositions of \citet{mendezabreu17}. (d) Visual Hubble type classification of our galaxies. (e) Stellar mass for bulges (red), discs (blue), and the whole galaxy (black) computed in this work.}
    \label{fig:properties}
\end{figure*}

Fig.~\ref{fig:properties} shows the distribution of the main spectro-photometric properties of the sample discussed in this paper. The individual values for each galaxy, and each component (bulge and disc), are provided in Table~\ref{tab:properties}. The stellar population properties will be provided in tabular form in a forthcoming paper. Throughout this paper, we use the photometric properties (mainly $r_{\rm e}$, $n$, and $B/T$) obtained from the multi-component photometric decompositions described in \citet{mendezabreu17}. They were performed using the Sloan Digital Sky Survey imaging \citep[SDSS-DR7;][]{abazajian09} in the $g-$, $r-$, and $i-$bands. The values used in this paper correspond to the $r-$band results, but the conclusions are not altered if we use any other band instead. The galaxy properties such as global effective radii and Hubble types are obtained from \citet{walcher14}. The stellar masses are derived from the stellar population analysis carried out applying {\sc Pipe3D} to the original CALIFA datacubes (global stellar mass) and individual component datacubes obtained from {\sc c2d} (bulge and disc stellar masses).

%---------------------------
%---------------------------

The typical effective radii for bulges are smaller than for discs, with those of the whole galaxies showing intermediate values (Fig.~\ref{fig:properties}, panel a). The effective radii for the disc component are computed from their fitted exponential scale-length such as $r_{\rm e}$ = 1.678$\times h$. Since we use broken profiles to describe the surface brightness profile of some galaxy discs in \citet{mendezabreu17}, it is worth mention that all values used in this work refer to the inner disc. We find mean values of 1.9, 11.5, and 5.3 kpc for bulges, discs, and galaxies, respectively. The distribution of the bulge S\'ersic index (panel b) is relatively constant for $n < 3$, dropping quickly for higher values. We obtain a mean S\'ersic index of 2.2 for our sample bulges. The $B/T$ luminosity ratio (panel c), described in detail in Sect.~\ref{sec:BT}, spans $0.02 < B/T < 0.73$ with a mean value of 0.26, therefore covering the whole range of possibilities from almost bulgeless galaxies to {\it nearly} ellipticals (a $B/T > 0.8$ is generally used to classify a galaxy as a single component elliptical; \citealt{mendezabreu18}). The wide range of bulges and discs covered in this study is also represented by their Hubble types (panel d). There is a drop for very late galaxies due to the lack of photometrically detected bulges in many of those galaxies. This is also reflected on the decreasing number of galaxies with low masses shown in panel d. The mean stellar masses (in logarithmic scale) for our bulges, discs, and whole galaxies are 10.0, 10.1, and 10.5 dex, respectively. Despite the limited number of galaxies in our sample, we discuss in Sect.~\ref{sec:massfunc} how it is statistically representative of the local Universe in the mass range $9.5 < \log{(M_{\star}/M_{\sun})} < 12$.

\begin{figure}
    \centering
    \includegraphics[bb=134 340 438 720,width=0.49\textwidth]{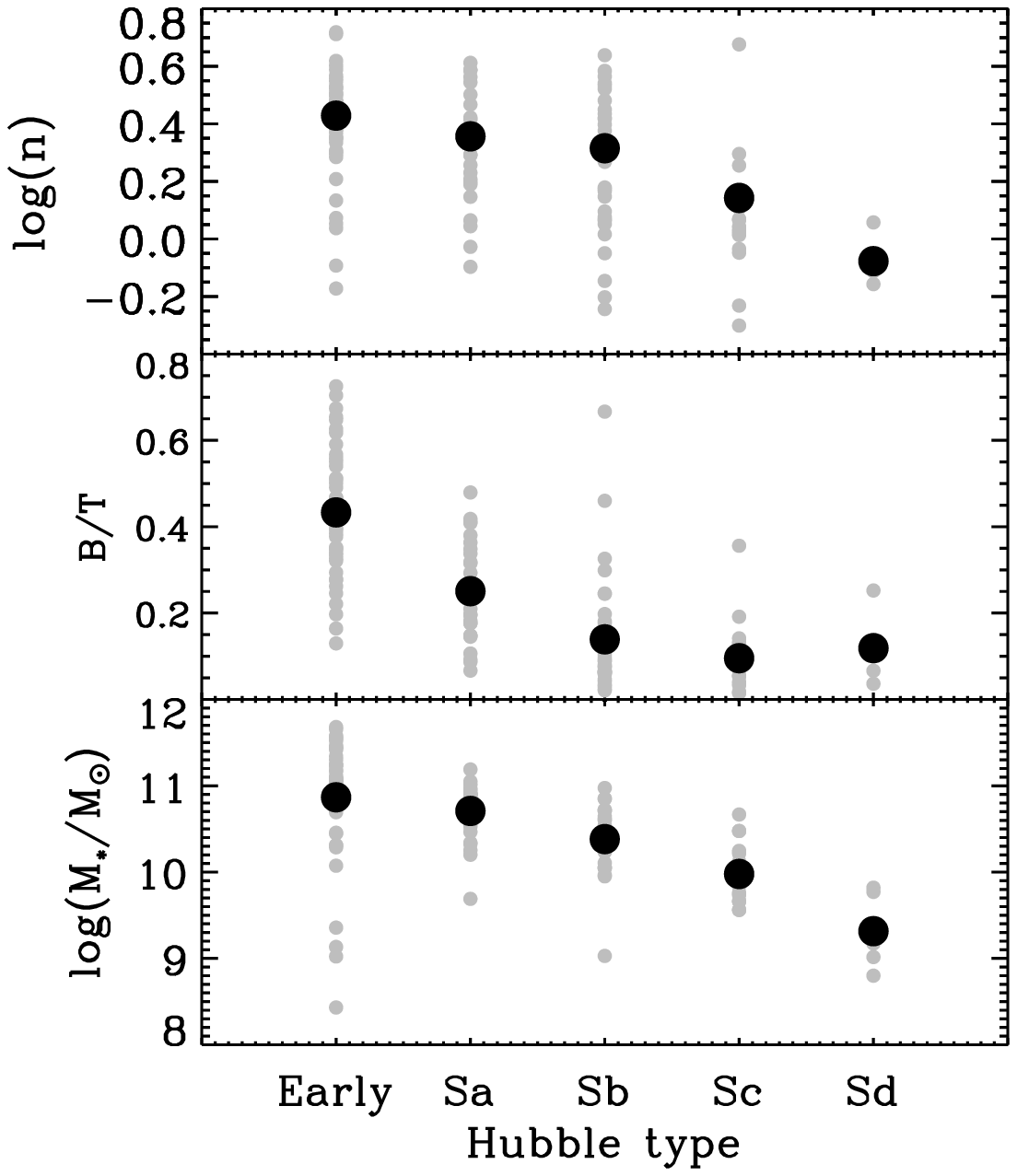}
    \caption{Distribution of the bulge S\'ersic index (top panel), $B/T$ luminosity ratio, and galaxy stellar mass (bottom panel) as a function of the Hubble type. Grey circles show the values for individual galaxies and black circles represent the mean values for each Hubble type.}
    \label{fig:phot_correlation}
\end{figure}

It is well known that some physical properties of bulges, discs, and (global) galaxies are correlated. In particular, the Hubble type correlates (with more or less scatter) with the S\'ersic index, $B/T$ luminosity ratio, and galaxy mass \citep{laurikainen07,weinzirl09,laurikainen10,mendezabreu17,gao19}. Fig.~\ref{fig:phot_correlation} shows these correlations for the sample used in this study. We note that, due to these correlations and for the sake of clarity, we will only show (and discuss) in this paper those relations with galaxy mass. Nonetheless, whenever we discuss trends with stellar mass they might be also interpreted as trends with Hubble type unless otherwise stated.

%---------------------------
%---------------------------

%---------------------------------------------------------------------------
\begin{table*}
 \centering
  \caption{Spectro-photometric properties of the sample}
  \label{tab:properties}
  \begin{tabular}{cccccccccc}
  \hline
Galaxy   & $r_{\rm e,b}$  & $r_{\rm e,d}$ & $r_{\rm e,gal}$ & $n$ & $B/T$ & HT & $\log{(M_{\star, \rm b}/M_{\sun})}$ & $\log{(M_{\star, \rm d}/M_{\sun})}$ & $\log{(M_{\star, \rm gal}/M_{\sun})}$  \\
 (1)  & (2) & (3) & (4) & (5) & (6) & (7) & (8) & (9) & (10) \\
\hline
\hline
IC0159  &  0.91  &  5.14  &  3.99  &  0.80  &  0.07  &  Sdm  &  8.93  &  9.68  &  9.77 \\
IC0208  &  0.52  &  6.41  &  4.25  &  0.57  &  0.02  &  Sbc  &  8.87  &  9.98  &  10.04 \\
IC0307  &  0.89  &  10.29 &  5.38  &  1.81  &  0.21  &  Sab  &  10.47 &  10.60 &  10.87 \\
IC0944  &  1.36  &  11.46 &  4.96  &  0.80  &  0.15  &  Sab  &  10.38 &  10.82 &  10.96 \\
IC1151  &  0.45  &  5.28  &  3.56  &  0.59  &  0.02  &  Scd  &  8.39  &  9.62  &  9.66 \\
\hline
\end{tabular}
\begin{minipage}{14cm}
(1) Galaxy name; (2), (3), (4) effective radius in the $r-$band for the bulge, disc (1.678$\times h$), and the galaxy in kpc; (5) S\'ersic index of the bulge; (6) bulge-to-total luminosity ratio in the $r-$band; (7) Hubble type obtained from \citet{walcher14}; (8), (9), (10) stellar mass computed using the stellar population analysis for the bulge, disc, and whole galaxy. We show only the first 5 galaxies of the sample, the remaining are available in the online version.
\end{minipage}\end{table*}
%---------------------------------------------------------------------------

%------------------------------------------------------
\subsection{Stellar mass functions of bulges and discs}
\label{sec:massfunc}

\begin{figure*}
    \centering
    \includegraphics[bb=50 380 520 720,width=0.99\textwidth]{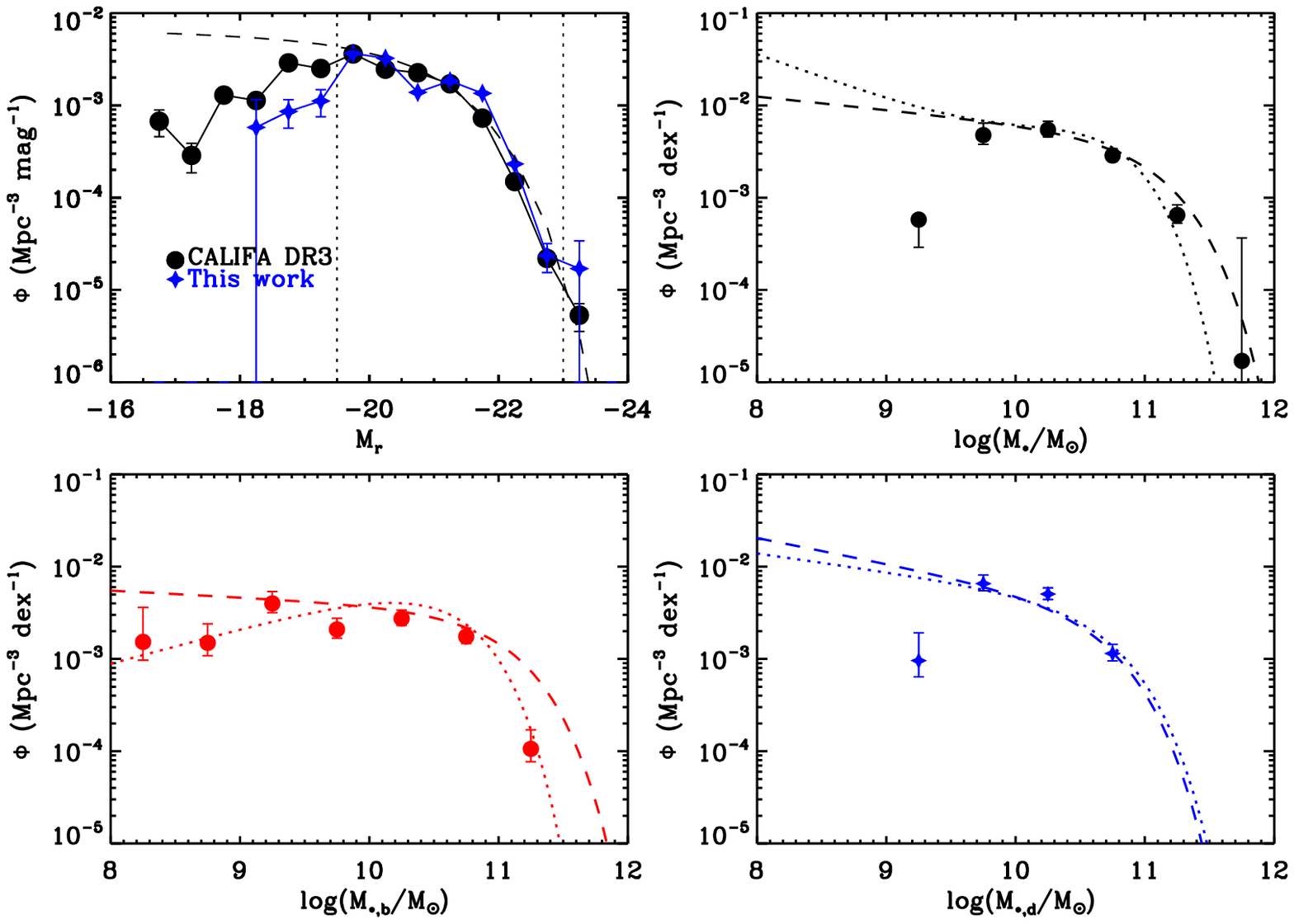}
    \caption{Top-left panel. Luminosity function for the CALIFA-DR3 sample of galaxies (blue circles) and a subsample of 105 galaxies from this work (blue stars). This subsample represents those galaxies extracted from the CALIFA mothersample (see text for details). The dashed line shows the SDSS luminosity function given by \citet{blanton03}. The vertical lines indicate the magnitude limits where our sample is complete. Top-right, bottom-left and bottom-right panels show the galaxy (black), bulge (red), and disc (blue) mass functions, respectively. The dashed lines show the comparison with the mass function derived by \citet{thanjavur16}. The dotted line in the galaxy mass function was obtained from \citet{kelvin14} and in the bulge and disc mass functions from \citet{moffett16}. Error bars show the Poissonian uncertanties.}
    \label{fig:mass-function}
\end{figure*}

The CALIFA-DR3 sample of galaxies is selected from the SDSS-DR7 based on an angular diameter and redshift selection, thus it is not strictly complete in either mass or volume. Nevertheless, \citet{walcher14} demonstrated that these constraints do not strongly bias the sample, and that it represents quite well the whole population of nearby galaxies. Therefore, they computed a volume correction for each individual galaxy that can be used to correct for the selection function. Therefore, although the final observed sample in the CALIFA-DR3 is not complete in volume, it is possible to reconstruct volume corrected sample properties using the CALIFA-DR3 \citep[see][]{sanchez16} within the completeness limits described in \citet{walcher14}, that is, $-19 > M_r > -23.1$. Fig.~\ref{fig:mass-function} shows the luminosity function in the $r-$band for both our sample of bulge-to-disc galaxies and the CALIFA-DR3. We note that volume corrections are not applicable for all galaxies in the CALIFA-DR3, but only for those in the CALIFA mothersample \citep[see][]{sanchez16}, so the luminosity function of our sample galaxies (Fig.~\ref{fig:mass-function}, top-left panel) includes only 105 galaxies. The remaining 24 galaxies were drawn from the so-called extended sample included in the CALIFA-DR3 \citep[see][]{sanchez16}. There is good agreement between the luminosity functions of CALIFA-DR3 and our sample, as well as for the SDSS luminosity function given by \citet{blanton03}, within the magnitude limits $-19.5 > M_r > -23$. This range is slightly shorter that the one found in \citet{walcher14}, but it demonstrates that our sample of bulge-to-disc galaxies is statistically representative of the whole nearby galaxies population within these limits. 

%---------------------------
%---------------------------

The stellar masses used in this study are obtained from the stellar population analysis described in Sect.~\ref{sec:c2d} and integrated over the whole FoV of CALIFA. In order to understand the mass limits where our sample is representative, we computed the stellar masses of those galaxies within the {\it complete} luminosity range $-19.5 > M_r > -23$. We find that they correspond to a lower limit in galaxy stellar mass  $\log{(M_{\star}/M_{\sun})} \simeq 9.5$. This is in agreement with the typical stellar mass where bulge-to-disc galaxies dominate over pure-disc systems in the CALIFA survey \citep[see Fig. 9 in][]{mendezabreu17}. Therefore, we consider that our sample is statistically representative of bulge-to-disc galaxies for masses with $\log{(M_{\star}/M_{\sun})} > 9.5$. Our sample consists of 121 galaxies within this stellar mass with 104 included in the original CALIFA mothersample. The galaxy, bulge, and disc stellar mass functions shown in Fig.~\ref{fig:mass-function} have been computed using the latter subsample. Due to the relative small number statistics of our sample we do not attempt to fit the stellar mass functions, but instead we compare them with those derived in other surveys such as GAMA \citep{moffett16} and SDSS \citep{thanjavur16}. These previous studies based their galaxy, bulge, and disc mass estimations on either a colour empirical calibration or SED fitting to broad-band imaging data, while our masses are derived from a full spectrum fitting technique to the CALIFA datacubes. However, we find a general good agreement with both works. Our galaxy mass function (Fig.~\ref{fig:mass-function}, top-right panel) does not help to solve the discrepancy between the previous works at low masses ($\log{(M_{\star}/M_{\sun})} < 9.5)$, however it seems to favour the Schechter modelling of \citet{thanjavur16} at the high mass end. This difference was discussed in \citet{moffett16} as due to the smaller volume covered by GAMA with respect to SDSS. Since the CALIFA sample is based on SDSS we consider this might also be the reason for our better match with \citet{thanjavur16}. On the contrary, our bulge mass function is better represented using the Schechter modelling of the GAMA data from \citet{moffett16}. They used a visual classification to separate single or multi-component galaxies, which is similar to our human-supervised multi-component decompositions performed in \citet{mendezabreu17}, and more accurate than the automatic classification  carried out by \citet{thanjavur16}. As previously stated, at stellar masses  $\log{(M_{\star}/M_{\sun})} < 9.5$ single component, pure-disc systems start to dominate the mass function and this might create the flat slope in the SDSS low-mass spheroids. The disc mass function is similar in both studies and close to our values except for our lowest mass bin.

As already indicated, the analysis of the luminosity and mass functions in this section suggest that our sample of bulge-to-disc galaxies is statistically representative of the whole population for galaxies in the magnitude range $-19.5 > M_r > -23$, which corresponds to galaxies in the stellar mass range $9.5 < \log{(M_{\star}/M_{\sun})} < 12$. Regarding the separated components, bulges and discs are well represented in the mass ranges $8 < \log{(M_{\star \rm,b}/M_{\sun})} < 11.5$ and $9.5 < \log{(M_{\star, \rm d}/M_{\sun})} < 11$, respectively. For the sake of clarity, we mark these limits whenever possible in forthcoming figures.

%---------------------------
\begin{figure}
    \centering
    \includegraphics[bb=50 380 530 720,width=0.49\textwidth]{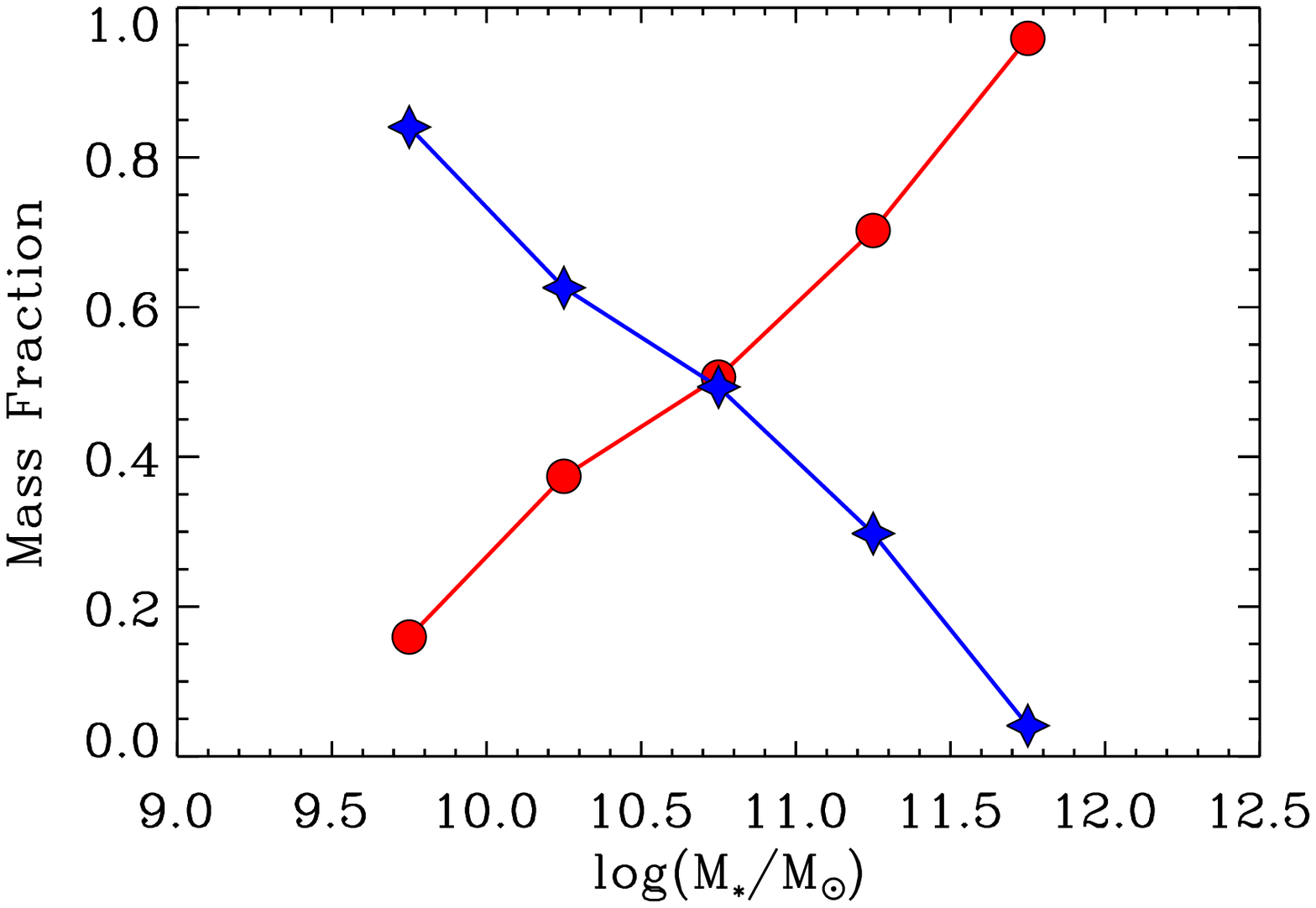}
    \caption{Volume corrected stellar-mass fraction of galactic bulges (red) and discs (blue) as a function of the global galaxy mass. Only the 104 galaxies in the mass range $9.5 < \log{(M_{\star}/M_{\sun})} < 12$ and belonging to the CALIFA mothersample are used in this analysis.}
    \label{fig:mass-fraction}
\end{figure}
%---------------------------

Using the previous global galaxy mass completeness limits ($9.5 < \log{(M_{\star}/M_{\sun})} < 12$) and the volume correction for each galaxy, we compute the fraction of mass in  bulges and discs for our sample. We find that 49\% and 51\% of the galaxy mass is in bulges and discs, respectively. These values are quite different from previous fractions reported in \citet{weinzirl09} and \citet{driver07}. The former used a sample of 143 bright spirals measuring that $\sim$70\% of the stellar mass is in discs, $\sim$10\% is in stellar bars and $\sim$20\% is in bulges. \citet{driver07} used
the Millennium Galaxy Catalog finding that 68.6\% of the stellar mass to be in discs, and 32.6\% in bulges. However, our results agree with the more recent study of \citet{moffett16} using the GAMA survey. They derived that 50\% of the local stellar mass density is in spheroids and 48\% in discs. However, they computed lower stellar mass densities in spheroids (1.24$\pm$0.49$\times$10$^8$ $M_{\sun}$ Mpc$^{-3}$) and discs (1.20$\pm$0.45$\times$10$^8$ $M_{\sun}$ Mpc$^{-3}$) compared to our 2.02$\times 10^8 M_{\sun}$ Mpc$^{-3}$ and 1.9$\times 10^8 M_{\sun}$ Mpc$^{-3}$ mass density for bulges and discs, respectively. We argue that most of the differences with previous studies rely on the different samples. On one side, we targeted only bulge-to-disc galaxies discarding barred systems. On the other hand the mass fraction of bulges and discs is strongly dependent on the mass range of the galaxy sample. Fig.~\ref{fig:mass-fraction} shows this trend for our sample using only those galaxies in the mass regime where our sample is complete and correcting for their available volume. We find a transition mass at $\log{(M_{\star}/M_{\sun})}$ $\sim$ 10.75, where high mass galaxies are mass dominated by the bulges and low mass galaxies by discs. This transition value is slightly lower than the one found in \citet[][$\log{(M_{\star}/M_{\sun})} \sim$ 10.9]{moffett16} and slightly larger than that measured by \citet[][$\log{(M_{\star}/M_{\sun})} \sim$ 10.5]{thanjavur16} but, taken into account our bin size (0.5 dex), is in good agreement with those works.

%------------------------------------------------------
%------------------------------------------------------
\subsection{B/T luminosity and mass ratio and its relation to galaxy mass}
\label{sec:BT}

\begin{figure}
    \centering
    \includegraphics[bb=50 380 530 720,width=0.49\textwidth]{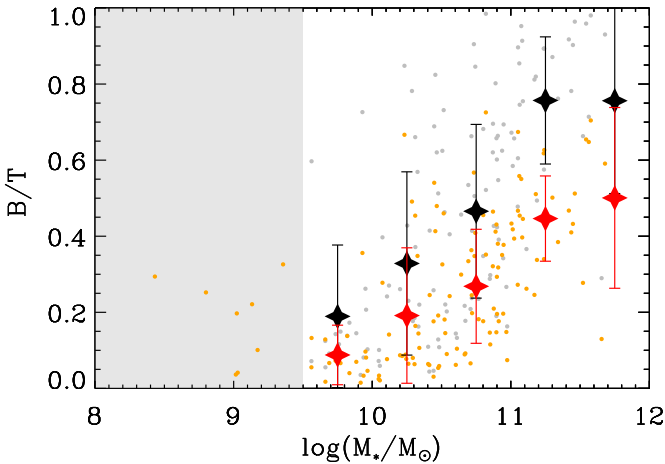}
    \caption{Distribution of the bulge-to-total ($B/T$) ratio as a function of galaxy mass. Small grey and orange circles show the results for individual galaxies using either mass or $r-$band light, respectively. Large black and red stars represent mean values using mass and $r-$band light, respectively. Errors bars show 1$\sigma$ deviations. Mean values are only computed in the complete global galaxy mass range ($9.5 < \log{(M_{\star}/M_{\sun})} < 12$). The shaded area shows the stellar mass range where our sample is incomplete.}
    \label{fig:BT_mass}
\end{figure}

Fig. \ref{fig:BT_mass} shows the $B/T$ distribution of our sample galaxies as a function of the global galaxy mass. We show the standard $B/T_{l}$ luminosity ratio obtained from the photometric decompositions performed in the SDSS $r-$band (orange circles and red stars; \citealt[][]{mendezabreu17}), but also the $B/T_{m}$ mass ratio (grey circles and black stars). The latter was derived using the results from our spectrophotometric decomposition, and computing the stellar mass of the bulges and discs through our stellar population analysis on the individual datacubes. The mean values are listed in Table \ref{tab:BT}.

%---------------------------------------------------------------------------
\begin{table}
 \centering
  \caption{Mean luminosity and mass $B/T$ ratios}
  \label{tab:BT}
  \begin{tabular}{ccc}
  \hline
$\log{(M_{\star}/M_{\sun})}$   & $B/T_l$  &  $B/T_m$  \\
 (1)  & (2) & (3)  \\
\hline
\hline
9.5  - 10.0 & 0.09$\pm$0.08  & 0.19$\pm$0.19 \\
10.0 - 10.5 & 0.19$\pm$0.18  & 0.33$\pm$0.24 \\
10.5 - 11.0 & 0.27$\pm$0.15  & 0.47$\pm$0.23 \\
11.0 - 11.5 & 0.45$\pm$0.11  & 0.76$\pm$0.17 \\
11.5 - 12.0 & 0.50$\pm$0.24  & 0.76$\pm$0.24 \\
\hline
\end{tabular}
\begin{minipage}{8cm}
(1) Mass interval in log units; (2) mean and standard deviation values of the $B/T$ luminosity ratio in the SDSS $r-$band; (3) mean and standard deviation values of the $B/T$ mass ratio. Only galaxies in the complete mass range have been used.
\end{minipage}\end{table}
%---------------------------------------------------------------------------

There is a clear trend between both $B/T_{l}$ and $B/T_{m}$ and the galaxy stellar mass, with more massive galaxies being more dominated by the bulge luminosity or stellar mass. This is consistent with the results shown in Fig.~\ref{fig:mass-fraction}, and with previous results in the literature \citep{weinzirl09, mendezabreu17}. We also show that on average  $B/T_{m} > B/T_{l}$ at all galaxy masses. This holds even when using the $B/T_{l}$ in the SDSS $i-$band. Therefore the simple hypothesis that, for a given galaxy, both the bulge and disc luminosities can be transformed into stellar masses assuming the same $M/L$ ratio is not accurate, otherwise $B/T_{m} \sim B/T_{l}$.

%---------------------------
%---------------------------

%---------------------------
\begin{figure*}
    \centering
    \includegraphics[bb=50 380 530 720,width=0.49\textwidth]{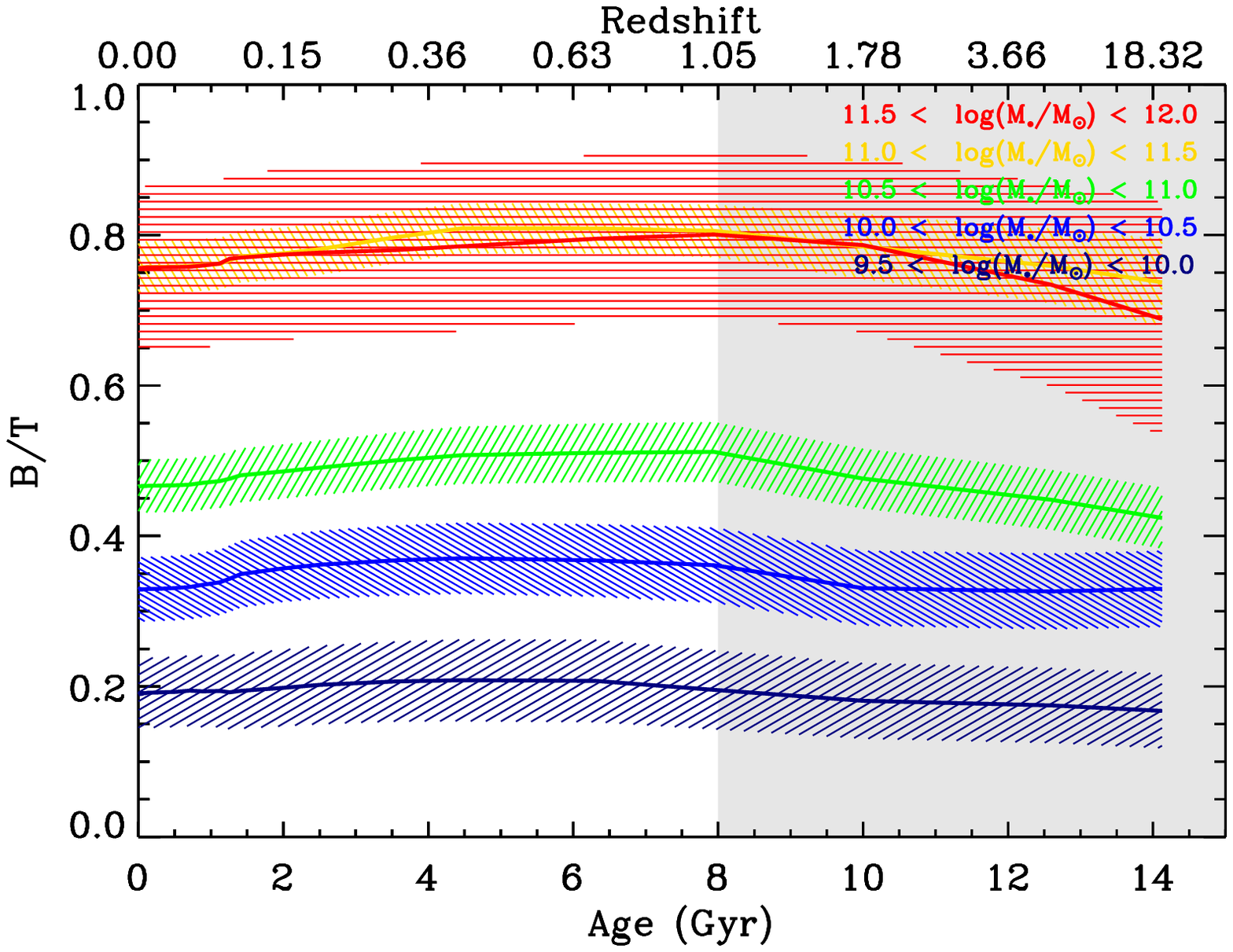}
    \includegraphics[bb=50 380 530 720,width=0.49\textwidth]{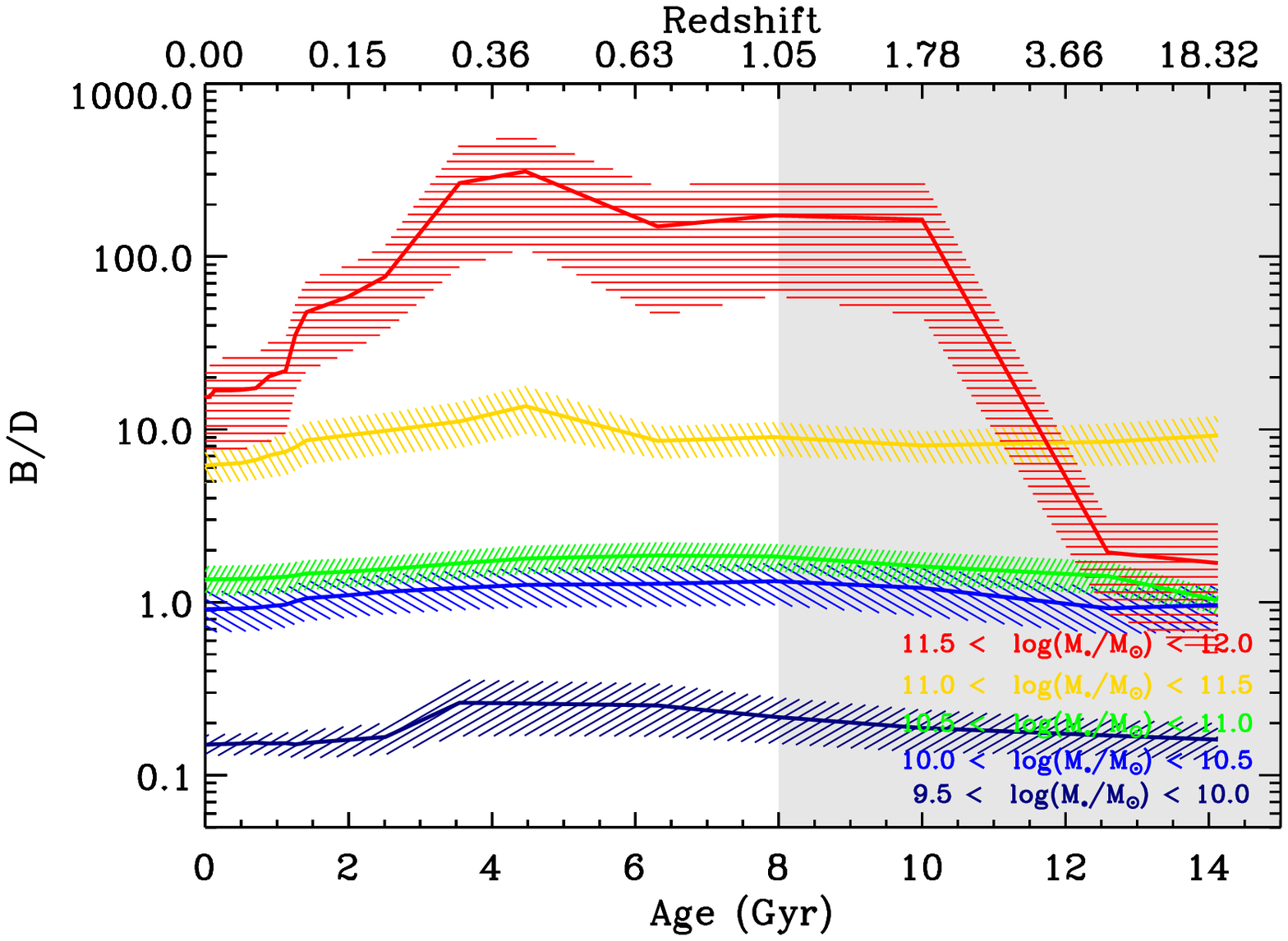}
    \caption{Time evolution of the bulge-to-total ($B/T$) mass ratio (left panel) and bulge-to-disc ($B/D$) mass ratio (right panel) for galaxies in different mass bins (see colors in the legend). The mass of every component has been computed using the stellar population analysis. The redshift is shown in the upper x-axis. The shaded area represents stellar ages (or redshifts) which are not well resolved in our stellar population analysis (see text for details).}
    \label{fig:BT_evolution}
\end{figure*}
%---------------------------

Fig.~\ref{fig:BT_evolution} shows the evolution of the $B/T_m$ and bulge-to-disc ($B/D_m$) mass ratio as a function of cosmic time and  redshift. We compute the amount of mass in each component (bulge and disc), for each time step, using their integrated star formation histories (SFH) across the FoV of CALIFA \citep[see][for details]{sanchez18}. This {\it fossil approach} has the caveat that our spectro-photometric definition of bulge and disc might be different at $z=0$ than at higher redshift, but it is otherwise very useful to provide a direct comparison with simulations. It is worth nothing that our definition does not take into account any difference between either in-situ and ex-situ formation of the stars in our bulges and disc or any stellar population mixing due to radial migration. It is also important to notice that the time resolution for stellar populations older than 8 Gyrs is scarce and therefore any evolution for $z \gtrsim 1$ should be taken carefully.  Numerical simulations show that the ex-situ stellar mass fraction in galaxies is only relevant at high masses \citep[$\log{(M_{\star}/M_{\sun})} > 11$;][]{rodriguezgomez17,tacchella19} and exclusively affect only the growth of bulges \citep{clauwens18}. Still, the dynamical redistribution of in-situ formed stars is not fully understood.

The evolution of the $B/T$ mass ratio with redshift is almost constant for a given galaxy stellar mass, showing typical values which are higher (at all redshifts) for more massive galaxies (similar to those shown in Fig. \ref{fig:BT_mass}). Therefore the relative growth of the bulge mass with respect to the global galaxy does not evolve significantly with time. Nonetheless, despite Fig.~\ref{fig:BT_evolution} showing a fairly constant behaviour of $B/T$ with redshift, we find that this parameter can become insensitive to variations when one of the components overly dominates. Indeed, the redshift evolution of the $B/D$ mass ratio (Fig.~\ref{fig:BT_evolution}, right panel) demonstrates that, for the most massive galaxies ($10.5 < \log{(M_{\star}/M_{\sun})} < 12$) and even if the $B/T$ does not change, the actual relative growth of bulges and disc can vary over orders of magnitude in mass.

The derived time evolution of the $B/T$ seems to be in contradiction with the evolution predicted in some the IllustrisTNG numerical simulations. \citet{tacchella19} found a significant evolution of the spheroid-to-total ($S/T$) mass ratio with redshift, in particular for their low-mass galaxies with $\log{(M_{\star}/M_{\sun})} < 10.5$. However, the evolution is milder in the EAGLE simulations \citep{clauwens18}. The convergent point in both simulations is that there is evidence for an epoch of disc formation for galaxies with $\log{(M_{\star}/M_{\sun})} > 10$. This is compatible with our $D/T$ evolution, which is also dependent on the mass and redshift. We further explore the mass growth of bulges and discs in a companion paper (M\'endez-Abreu et al. in preparation).

%------------------------------------------------------
%------------------------------------------------------
\subsection{Mass-size relations for bulges and discs}
\label{sec:mass-size}

\begin{figure*}
    \centering
    \includegraphics[bb=100 370 590 720,width=0.45\textwidth]{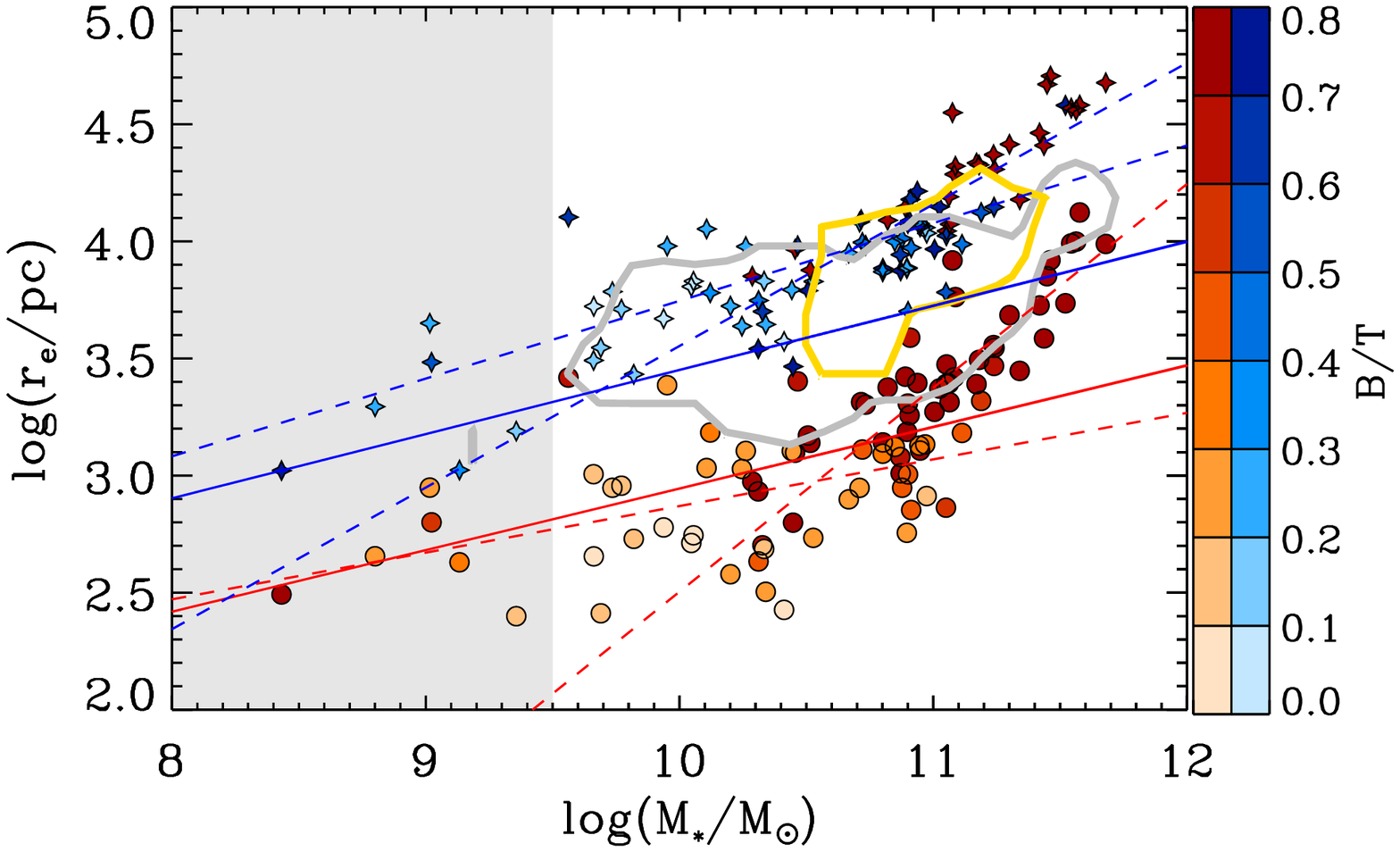}
    \includegraphics[bb=45 370 550 720,width=0.45\textwidth]{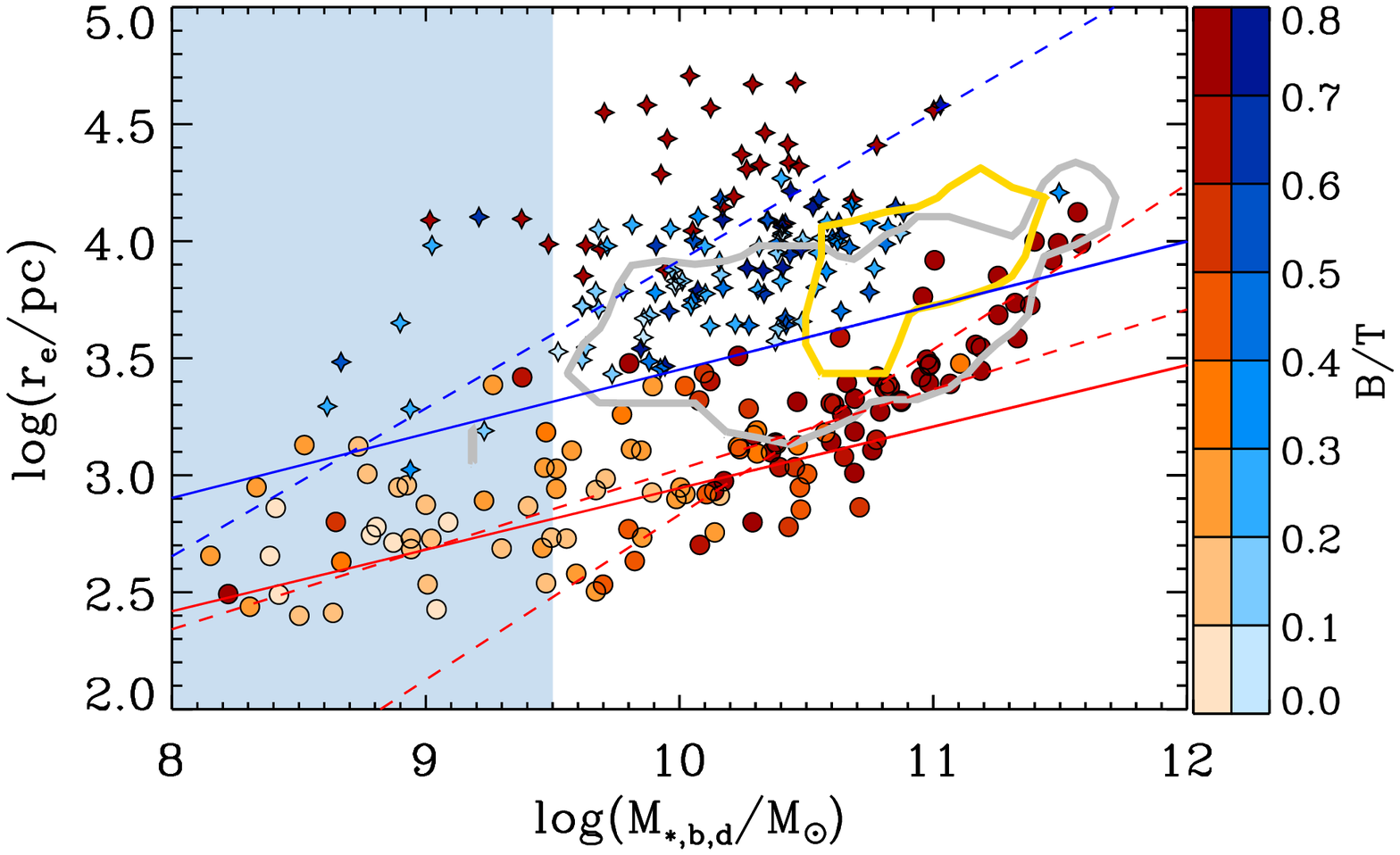}
    \caption{Mass-size relation for the bulges (circles) and discs (stars) of our galaxy sample. The left and right panels show the distributions as a function of the global galaxy mass and the mass of each component, respectively. All stellar masses are derived from our stellar population analysis. The effective radii of both bulges and discs (1.678$\times h$) are obtained from the $r$-band photometric decompositions of \citet{mendezabreu17}. Grey isocontours represent the mass-size relation of our galaxy sample without separating their structures. Golden isocontours show the mass-size relation of the photometrically defined {\it pure} ellipticals in the CALIFA sample \citep[see][for details]{mendezabreu18}. The $B/T$ luminosity ratio for each galaxy is coloured in reddish (bulge) and bluish (disc) colors according to the colorbars. The blue and red solid lines show the best-fit obtained by \citet{lange16} for their sample of discs and spheroids, respectively. The blue and red dotted lines represent the best-fits (see Table~\ref{tab:mass-size}) obtained for our sample of discs and bulges, respectively. To perform our fit to the mass-size relation we divided the sample into two mass bins: $8 < \log{(M_{\star}/M_{\sun})} < 10.5$ and $10.5 < \log{(M_{\star}/M_{\sun})} < 12$ to capture the clearly appreciated change in the slope. The grey and blue shaded areas show the global and disc stellar mass range where our sample is incomplete, respectively.}
    \label{fig:mass-size}
\end{figure*}

Fig.~\ref{fig:mass-size} shows the mass-size relation for the bulges and discs in our galaxy sample. The size of the galaxy components is obtained from the $r-$band photometric decompositions of \citet{mendezabreu17}. The bulge effective radius ($r_{\rm e}$), the disc effective radius (1.678$\times h$), and the global galaxy effective radius ($r_{\rm e, gal}$ obtained from a single S\'ersic fit) are therefore used as measure of the galaxy/component size. The stellar masses of the bulges, discs, and for the global galaxies are computed from the stellar population analysis of their associated CALIFA datacubes. Fig.~\ref{fig:mass-size} also shows the mass-size relation for the galaxies in our sample as if they were considered a single component (grey isocontours) and for a comparison sample of photometrically defined 'pure' elliptical galaxies (golden contours, see Sect~\ref{sec:sample}).

%--------------------------- 
%---------------------------

We first focus on the global galaxy stellar mass vs. component size (left panel in Fig.~\ref{fig:mass-size}). We find that, for a given galaxy mass, discs are always larger than bulges at all masses ranging from $8 < \log{(M_{\star}/M_{\sun})} < 12$. This is consistent with the general picture that spheroids and discy galaxies show distinct trends \citep{kauffmann03}. Both 'pure' elliptical galaxies and complete galaxies (i.e., without any decomposition) show an intermediate behaviour between bulges and discs. The different behaviour of ellipticals and bulges has also been pointed out in previous studies \citep{gadotti09,laurikainen10}. Our global galaxy stellar mass vs. component size relations show an upturn at $\log{(M_{\star}/M_{\sun})} \sim 10.5-11$ for all the systems used in this study: bulges, discs, ellipticals (even if they only cover the massive end), and global galaxies. To describe the shape of the relation we perform a fit to the $M_{\star}$-$r_{\rm e}$ using a simple power law \citep{shen03,lange16}. However, we find this simple modelling is not a good representation of the relations due to the aforementioned upturn. Therefore, we divide the sample into two mass bins: $8 < \log{(M_{\star}/M_{\sun})} < 10.5$ and $10.5 < \log{(M_{\star}/M_{\sun})} < 12$ and perform two different fits to capture the change in the slope. The best fit values are shown in Table \ref{tab:mass-size} and they are also shown in Fig.~\ref{fig:mass-size}. For the sake of comparison we also plot the results from \citet{lange16} using their sample of $z=0$ discs and spheroids analysed in the $r-$band for galaxies with $\log{(M_{\star}/M_{\sun})} > 9$. We find a good agreement in the slopes of the mass-size relation for bulges and discs when using our low-mass galaxy sample. The zero-point of the relation for the bulges is also in good agreement, but our discs are systematically larger than those analysed in \citet{lange16}. In addition, previous studies of the $M_{\star}$- $r_{\rm e}$ have found that, when divided into components, they are typically less curved than the global galaxy relation \citep{lange16,bernardi14} except for the discs in late-type galaxies, that cannot be fit with a single power law. This is in contradiction with our results since we find the upturn in all the systems analysed in this study.

%---------------------------------------------------------------------------
\begin{table}
 \centering
  \caption{Best-fit values to the $M_{\star}$-$r_{\rm e}$ relation}
  \label{tab:mass-size}
  \begin{tabular}{lcc}
  \hline
$\log{(M_{\star}/M_{\sun})}$   &  $a$ & $b$   \\
 (1)  & (2) & (3)  \\
\hline
\hline
Bulge 8.0  - 10.5 & 0.88$\pm$0.48   & 0.20$\pm$0.05 \\
Bulge 10.5 - 12.0 & -6.2$\pm$0.39   & 0.87$\pm$0.04 \\
Disc 8.0 - 10.5   & 0.43$\pm$0.41   & 0.33$\pm$0.04 \\
Disc 10.5 - 12.0  & -2.49$\pm$0.33  & 0.61$\pm$0.03 \\
\hline
Bulge 8.0  - 10.5 & -0.39$\pm$0.37   & 0.34$\pm$0.04 \\
Bulge 10.5 - 12.0 & -4.22$\pm$0.39   & 0.71$\pm$0.04 \\
Disc 8.0 - 10.5   & -2.4$\pm$0.60   & 0.63$\pm$0.06 \\
\hline
\end{tabular}
\begin{minipage}{8cm}
(1) mass interval in log units used for the bulge and disc fits; (2) and (3) $a$ and $b$ best fitted coefficients to the relation $\log{r_{\rm e}} = a + b \log{(M_{\star}/M_{\sun})}$, respectively. Above and below the line show results for global galaxy mass and individual component mass, respectively
\end{minipage}\end{table}
%---------------------------------------------------------------------------

Another piece of information that can be obtained from analysing the mass-size relation using the stellar masses of the individual components (bulges and discs). This is shown in the right panel of Fig.~\ref{fig:mass-size}. The best fit values to the power law obtained for the bulge distribution show a similar behaviour to those derived using the global galaxy mass, even if the actual values of both slopes and zero-points are different. In addition, we find that the scatter around the best fit is lower when considering only the mass of the bulge component than with the global galaxy mass. The situation is completely different for galaxy discs where i) there are only a few systems with $\log{(M_{\star,\rm d}/M_{\sun})} > 10.5$ and ii) both the best-fit values and the scatter are less constrained than when using the global galaxy mass.

Fig.~\ref{fig:mass-size} also includes colour-coded information about the $B/T$ luminosity ratio for each galaxy. The global galaxy stellar mass vs. component size (left panel) shows the trend between $B/T$ and stellar mass discussed in Sect.~\ref{sec:generalprop}, where more massive galaxies host more prominent bulges. We find that the upturn in the mass-size relation happens at $\log{(M_{\star}/M_{\sun})} \sim 10.5$ and $B/T \sim 0.2$. This result holds for bulges when using only their stellar mass (right panel), but not for discs.

%------------------------------------------------------
\subsubsection{Mass-size for bulges and discs and the Hubble type}

The mass-size relation of bulges and discs as a function of the Hubble type follows a similar trend as with the $B/T$ mass ratio shown in Fig.~\ref{fig:mass-size}. Galaxies with earlier Hubble types are generally more massive and they have larger bulges and discs with respect to later types. This is expected due to the known relations between the Hubble types, $B/T$, and S\'ersic index shown in Sect.~\ref{sec:generalprop}. Due to the limited size of our sample, and their low number of very late-type galaxies, we cannot create individual mass-size relations for different Hubble types. Nevertheless, Table~\ref{tab:sizes} shows the median sizes of our bulges and discs in different bins of mass and Hubble type (only for bins with more than two galaxies).

We find that, independently of the Hubble type of the galaxy, bulges and discs are always larger in size as the stellar mass of the galaxy increases. Therefore the mass-size relation holds for any Hubble type. However, for a fixed galaxy mass bin we do not find a statistically significant difference in either bulges or discs sizes, except for the lowest mass bin ($9 < \log{(M_{\star}/M_{\sun})} < 10$) where the effective radii of disc increases towards later Hubble types.

%---------------------------------------------------------------------------
\begin{table}
 \centering
  \caption{Effective radii of bulges and discs}
  \label{tab:sizes}
  \begin{tabular}{lccc}
  \hline
HT   & $9 - 10$ & $10 - 11$ & $11 - 12$\\
 (1)  & (2) & (3) & (4)  \\
\hline
\hline
TOTAL (B)   & 0.7$\pm$0.8  & 1.1$\pm$0.8   & 3.1$\pm$3.1       \\
TOTAL (D)   & 4.8$\pm$3.6  & 8.8$\pm$4.1   & 21.0$\pm$12.7     \\
TOTAL (B/D) & 0.15         & 0.12          & 0.14              \\
Early (B)   & 0.4$\pm$0.2  & 1.1$\pm$0.6   & 3.5$\pm$3.1       \\
Early (D)   & 1.5$\pm$1.0  & 7.5$\pm$3.6   & 21.2$\pm$12.5     \\
Early (B/D) & 0.28         & 0.14          & 0.17              \\
Sa    (B)   & -            & 1.0$\pm$0.9   & 1.4$\pm$0.7       \\
Sa    (D)   & -            & 9.6$\pm$5.3   & 12.0$\pm$3.9      \\
Sa    (B/D) & -            & 0.11          & 0.12              \\
Sb    (B)   & 0.5$\pm$1.2  & 1.0$\pm$0.8   & -                 \\
Sb    (D)   & 4.8$\pm$3.9  & 9.5$\pm$3.4   & -                 \\
Sb    (B/D) & 0.10         & 0.10          & -                 \\
Sc    (B)   & 0.8$\pm$0.9  & 1.5$\pm$0.6   & -                 \\
Sc    (D)   & 6.1$\pm$3.7  & 8.9$\pm$3.0   & -                 \\
Sc    (B/D) & 0.13         & 0.17          & -                 \\
Sd    (B)   & 0.9$\pm$0.3  & -             & -                 \\
Sd    (D)   & 5.1$\pm$2.9  & -             & -                 \\
Sd    (B/D) & 0.18         & -             & -                 \\
\hline
\end{tabular}
\begin{minipage}{8cm}
(1) Hubble type: TOTAL includes all galaxies in the sample, Early encompasses visually classified ellipticals that could host an extended disc (see Sect.~\ref{sec:sample} for details) and lenticulars; (2), (3), (4), and (5) are the global galaxy mass intervals used to compute the medians and 1$\sigma$ distributions in units of $\log{(M_{\star}/M_{\sun})}$. The effective radii are given in units of kpc. The B, D, and B/D separate the effective radii for bulges, discs, and their ratio, respectively. Only bins with more than two galaxies are shown. 
\end{minipage}\end{table}
%---------------------------------------------------------------------------

%------------------------------------------------------
\subsubsection{Mass-size evolution for bulges and discs}

\begin{figure*}
    \centering
    \includegraphics[bb=50 340 530 710,width=0.9\textwidth]{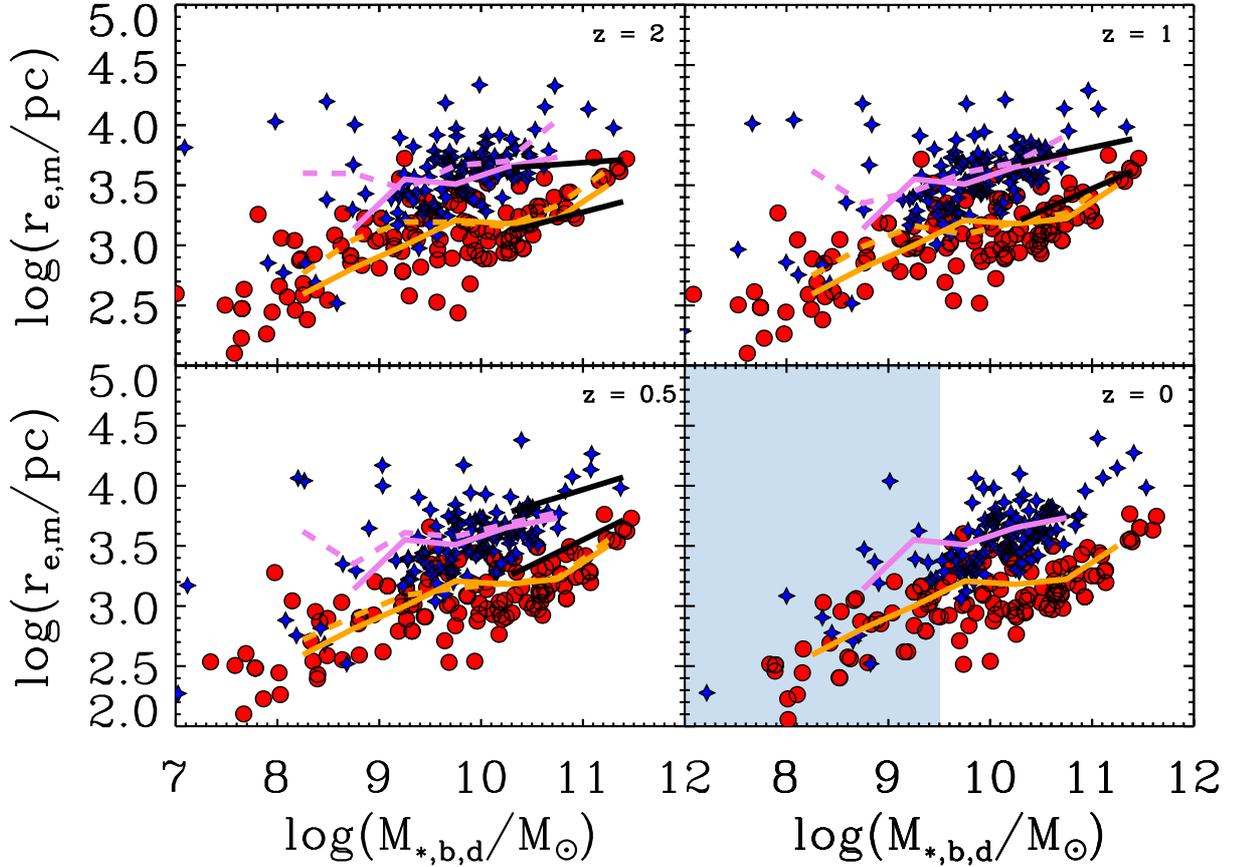}
    \caption{Mass-size relation for the bulges (circles) and discs (stars) of our sample galaxies at different redshifts ($z=2$, upper left; $z=1$, upper right; $z=0.5$, bottom left; $z=0$, bottom right). The stellar mass and effective radius correspond to the one for each component. At $z > 0$, different panels show the mass of each component at the corresponding redshift as computed from the fossil record method. The effective radii are computed using the reconstructed mass surface density at different ages (redshift) from their stellar populations, therefore they are half mass radii ($r_{\rm e,m}$)
    The orange and violet solid lines in all panels show the mean values of the $z=0$ relation computed in bins of 0.5 magnitude for bulges and discs (only bins with more than 5 galaxies are shown), respectively. The dashed lines show the mean values at the corresponding redshift. 
    Black lines show the best fit obtained from \citet{dimauro19} to their sample of $\log{(M_{\star}/M_{\sun})} > 10.3$ bulges and discs at high redshift.The blue shaded area shows the disc stellar mass range where our sample is incomplete.}
    \label{fig:mass-size-evolution}
\end{figure*}

Fig.~\ref{fig:mass-size-evolution} shows the redshift evolution of the mass-size relation for the bulges and discs in our sample. To this aim, we compute the mass of each component at three different redshifts ($z=2, 1,$ and $0.5$) using the derived SFHs for each galaxy/component, and compare it with the results at $z=0$ (orange and violet solid lines). The size evolution of the different components is obtained using the reconstruction of the mass surface density at different ages (redshift) of their stellar populations \citep[see][for an example]{martinnavarro19}. The limited spatial resolution of the CALIFA data does not allow us to perform a proper photometric decomposition onto these reconstructed images, therefore, we measure the effective radii of both bulges and discs as the radii enclosing half of the total mass for each component. We notice here that the sizes used in Fig.~\ref{fig:mass-size-evolution} are therefore half-mass radii ($r_{\rm e,m}$) and the $z=0$ relation is thus different from the one shown in Fig.~\ref{fig:mass-size}.

%---------------------------
%---------------------------

Fig.~\ref{fig:mass-size-evolution} suggests that discs are always (at all redshifts) larger than bulges for the same stellar mass. This is consistent with previous measurements in the literature using samples at different redshift \citep{bruce14,dimauro19}. Despite the uncertainties inherent to our method, the mass-size relation of bulges is already well-defined at $z \sim 2$, showing even the upturn at high masses. In fact, there is little evidence for a mass evolution of the upturn with redshift. This is expected since the mass evolution of bulges is negligible in this redshift range, $\sim 1.3$ (most of the stars in bulges were formed at $Ages > 10$ Gyrs). The mass evolution of discs is stronger than for bulges (see blue stars in Fig.~\ref{fig:mass-size-evolution}), with differences being dependent on galaxy mass. Low mass discs ($\log{(M_{\star, \rm d}/M_{\sun})} < 10$) show the larger mass growth, a factor $2.4$ since redshift $z=2$. We find a negligible size evolution of both components at all masses. This is contrary to previous results found in the literature. \citet{bruce14} used a sample of massive galaxies ($\log{(M_{\star}/M_{\sun})} > 11$) from CANDELS UDS and COSMOS fields to obtain that the median sizes of the bulge and disc components increased by a factor of 3.09 and 1.77 since redshift $1 <z < 3$, respectively. This is in agreement with the values we derive from  \citet{vanderwel14}. Other observational results have also reported a stronger size evolution of massive bulges with respect to massive disc using two-component photometric decompositions at high redshift \citep{bruce12,lang14}. Two reasons can be argued for this discrepancy: i) as explained before, we are showing the evolution of the half mass radius ($r_{\rm e,m}$) for both components while all previous results are based on the half light radius. Therefore, differences in the stellar populations, morphological k-corrections, or tracing the same galaxies with redshift might be biasing the results; ii) both internal or external dynamical processes such as radial migration of stars and gas, minor mergers of galaxies, or adiabatic expansion, which are not accounted for in our analysis, are predominant in the size evolution of bulges and discs. Fig.~\ref{fig:mass-size-evolution} also shows the comparison with recent results from \citet{dimauro19}. They studied the mass-size relation of bulges and discs at high redshift for massive systems ($\log{(M_{\star}/M_{\sun})} > 10.3$). They also show a large size evolution of discs and, considering the previous caveats, the milder size evolution for their bulges is qualitatively in agreement with our results.

%------------------------------------------------------
%------------------------------------------------------
\subsection{Sersic index vs. mass}

\begin{figure}
    \centering
    \includegraphics[bb=70 380 530 710,width=0.49\textwidth]{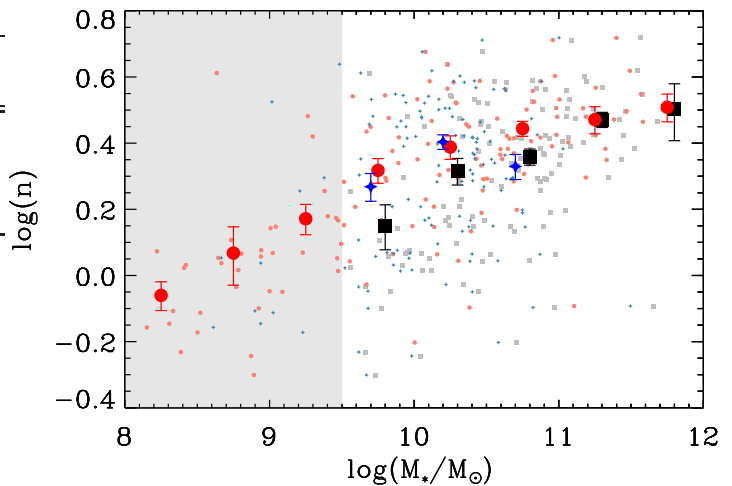}
    \caption{Distribution of the bulge S\'ersic index as a function of the global galaxy mass (grey small squares), bulge mass (small salmon circles), and disc mass (small cyan stars) for our sample galaxies. Large black squares, red circles, and blue stars represent the mean values in global galaxy, bulge, and disc masses, respectively. The mean values are only represented for bins where our sample is statistically representative (see Sect.~\ref{sec:massfunc}). Blue stars and black squares have been slightly shifted from the center of the bin for visualization purposes. The grey shaded area shows the global stellar mass range where our sample is incomplete.}
    \label{fig:n}
\end{figure}

Fig.~\ref{fig:n} shows the distribution of the bulge S\'ersic index as a function of the galaxy, bulge, and disc stellar mass. We also computed the mean $n$ values in bins of mass for the different components to highlight possible trends. These binned values were only computed in mass intervals where our sample can be considered statistically representative (see Sect.~\ref{sec:massfunc}). We find a clear positive trend between S\'ersic index and both the bulge and galaxy stellar mass. Thus, more massive galaxies, and galaxies with more massive bulges, show larger S\'ersic indices than the lower mass ones. We also find a steeper slope in the relation for galaxy mass with respect to the bulge mass. The trend between bulge S\'ersic index and galaxy luminosity or mass has already been studied in the literature \citep{weinzirl09, laurikainen10} and it is also related to the $B/T$ relation with mass (see Sect.~\ref{sec:generalprop}). Table \ref{tab:sersic} shows the fraction of massive galaxies ($\log{(M_{\star}/M_{\sun})} > 10$) with different S\'ersic index in our sample compared with \citet{weinzirl09}. We also compute those values corrected by volume as described in Sect.~\ref{sec:massfunc}. We find significant differences between our study and \citet{weinzirl09}. Despite this, both works find a low fraction of bulges with $n \geq 4$, most of our bulges are described by a S\'ersic index in the range $2 < n < 4$, whereas most of \citet{weinzirl09} bulges have $n \leq 2$. One possible explanation might be that we are specifically discarding barred galaxies in our study, which could host lower S\'ersic bulges than non-barred ones, whereas \citet{weinzirl09} fractions do include barred systems. However, they also separate barred and non-barred galaxies in the photometric decompositions and they did not find a significant difference in the S\'ersic index vs. galaxy mass relation for both types of galaxies.

We do not find a correlation between the S\'ersic index and the stellar mass of the disc. However, we notice that our range of disc masses where the sample is statistically representative is relatively narrow. In any case this might indicate that the concentration of bulges do not depend on their surrounding disc, but only on the galaxy and bulge mass. Numerical simulations with enough mass resolution have attempted to reproduce the $n$ vs. stellar mass relation. \citet{du20} using IllustrisTNG  (TNG100) simulations found that the S\'ersic index of their bulge components increases from $n$ = 0.9$^{+0.7}_{0.4}$ for $\log{(M_{\star}/M_{\sun})}$ = 10 galaxies to $n$ = 1.4$^{+0.6}_{-0.4}$ in galaxies of $\log{(M_{\star}/M_{\sun})}$ = 11. These values are smaller than those measured with our observations, but the authors already discussed that TNG100 generates systematically less compact spheroids than observed. Other simulations have also found that bulges become more compact (using other concentration parameters such as $C_{82}$) at higher masses \citep{tacchella19}, but a realistic comparison with the S\'ersic index remains to be done.

%---------------------------------------------------------------------------
\begin{table}
 \centering
  \caption{S\'ersic index in massive bulge-to-disc galaxies}
  \label{tab:sersic}
  \begin{tabular}{lccc}
  \hline
  & This work & This work (VC) & W09\\
   & (1) & (2) & (3) \\
\hline
\hline
Bulges with $n \geq 4$  & 5.8\%  & 6.8\%  & 1.8$\pm$1.2\% \\
Bulges with $2 < n < 4$ & 57.7\% & 67.5\% & 23.9$\pm$4.01\% \\
Bulges with $n \leq 2$  & 36.5\% & 25.7\% & 74.3$\pm$4.11\% \\
Bulges with $n \geq 2$  & 63.5\% & 74.2\% & 25.7$\pm$4.11\% \\
\hline
\end{tabular}
\begin{minipage}{8cm}
(1) Fraction of galaxies with $\log{(M_{\star}/M_{\sun})} > 10$ and bulges within the various S\'ersic indices constraints as measured in this work; (2) similar to (1), but using volume corrected fractions; (3) fraction of galaxies with $\log{(M_{\star}/M_{\sun})} > 10$ and bulges within the various S\'ersic index constraints as measured by \citet{weinzirl09}.
\end{minipage}\end{table}
%-----------------------------------------------------------------------

%---------------------------
%---------------------------

%------------------------------------------------------
%------------------------------------------------------
\section{Discussion}
\label{sec:discussions}
In this section we discuss our previous results in the context of both the structures and galaxy morphological evolution. It is worth reminding here that our sample does not include barred galaxies. Bars are recognised as the main internal mechanism driving secular evolution in galaxies and their effects on the evolution of galaxy morphology and mass growth will be studied in a forthcoming work. Therefore, our results are representative of the Hubble sequence branch of non-barred galaxies.

%------------------------------------------------------
\subsection{The rise of galactic bulges}

The results presented throughout this paper show a tight relation between the properties of bulges and their host galaxies. This connection has been shown, and used, in the literature to study different aspects of galaxy evolution, for instance to relate the central supermassive black hole properties with the properties of their galaxies and bulges \citep{magorrian98,kormendyho13}. However, this link has been better understood in the context of very massive galaxies, where bulges contribute to most of the galaxy mass and their origin is generally associated to high redshift mergers. However, we find that bulge and galaxy properties are closely related at all galaxy masses (at least in the mass range where we are statistically representative 9.5 < $\log{(M_{\star}/M_{\sun})}$ < 11). This can be seen in the lack of evolution of the $B/T$ mass ratio with redshift (Fig.~\ref{fig:BT_mass}). Despite the fact that more massive galaxies host more massive bulges (Fig.~\ref{fig:BT_mass}), we find a lack of evolution at all galaxy masses. Thus, massive galaxies hosting massive bulges show large values of $B/T$ at all time whereas low-mass galaxies behave similarly over time, but with lower $B/T$ values. 

Numerical simulations provide different views on this aspect. \citet{avilareese14}, using a semi-empirical approach, found that massive galaxies, $\log{(M_{\star}/M_{\sun})}$ > 10.5, do set-up their $B/T$ at high redshift with little variation through time. However, their low-mass galaxies at  $z=0$, i.e., those within the mass range 9 < $\log{(M_{\star}/M_{\sun})}$ < 10.5, that have low $B/T$ values, do present larger $B/T$ values in the past since they are mostly modified by the smooth growth of discs since $z\sim1$. In fact, we also find a different behaviour for the growth of disc, where they mostly influence the galaxy morphology for high-mass galaxies ($\log{(M_{\star}/M_{\sun})}$ > 10.5, see next section). Using hydrodynamical simulations, \citet{tacchella19} also show a decrease of the spheroid-to-total $S/T$ mass fraction with time for low-mass galaxies which is only mild at high masses whereas \citet{clauwens18} find a similar $B/T$ vs. mass evolution at all redshifts. At this point, it is worth noting that both \citet{tacchella19} and \citet{clauwens18} use a kinematic definition for their spheroid that produces a 'inverted' bell-shape $S/T$ distribution with galaxy mass. It is known that this definition produces important differences with respect to the typical photometric definition obtained from observations, particularly in the low-mass regime where galaxies are photometrically considered as one component, but they are kinematically hot (see \citealt{sanchezjanssen10} for an observational description of this phenomenon). Indeed, if we use their mass concentration parameters, which show a more similar monotonically increasing profile with galaxy mass to our $B/T$ definition, they do not find a significant evolution with redshift. As mentioned before, a caveat associated with our analysis is that structure evolution beyond $z>1$ must be taken carefully due to the scarce resolution of the stellar population models. Besides the aforementioned differences, a picture where the bulge component and the global galaxy are intimately linked at all stellar masses arises in both observation and simulations.

Fig.~\ref{fig:mass-size} also reflects the similarities between bulges and whole galaxies. Indeed, the mass-size relation for galactic bulges shows the high-mass upturn typical of high-mass galaxies. Again, this might be understood as massive galaxies being mainly 'bulges' ($B/T \sim 0.8$), however, galaxy bulges also show a flattening towards low-mass systems which is also observed in global galaxies. There is obviously an offset in size (and mass) between both systems, but it is significant that while bulges at all masses follow the mass-size relation, galactic discs do not (see next section). Finally, the concentration of galaxy bulges, as estimated from their S\'ersic index, also shows a monotonic relation with both galaxy and bulge stellar mass (Fig.~\ref{fig:n}). There is a different slope, which is more likely related to the offset in mass, but it is again clear the tight dependence in the evolution of galaxies and bulges even in low-mass systems ($\log{(M_{\star}/M_{\sun})}$ > 9.5 for galaxies and $\log{(M_{\star, \rm b}/M_{\sun})}$ > 8 for bulges).

%------------------------------------------------------
\subsection{The puzzle of galactic discs}

Galactic discs in our sample show a different behaviour than bulges in their relation with global galaxy properties. Despite our sample is statistically complete in the galaxy mass range 9.5 < $\log{(M_{\star}/M_{\sun})}$ < 12, and the bulges hosted by these galaxies span a mass range 8 < $\log{(M_{\star, \rm b}/M_{\sun})}$ < 11.5, galaxy discs only appear in a much narrower mass range 9.5 < $\log{(M_{\star, \rm d}/M_{\sun})}$ < 11. This leads to the idea that the efficiency of galaxies to form discs is much more limited than in the case of bulges, and therefore their relation with global galaxy properties is also different. We will further explore the efficiency of disc formation in M\'endez-Abreu et al. (2020, in preparation) and here we will focus on the relation with galaxy morphological properties.

\begin{figure}
    \centering
    \includegraphics[bb=80 380 560 720,width=0.45\textwidth]{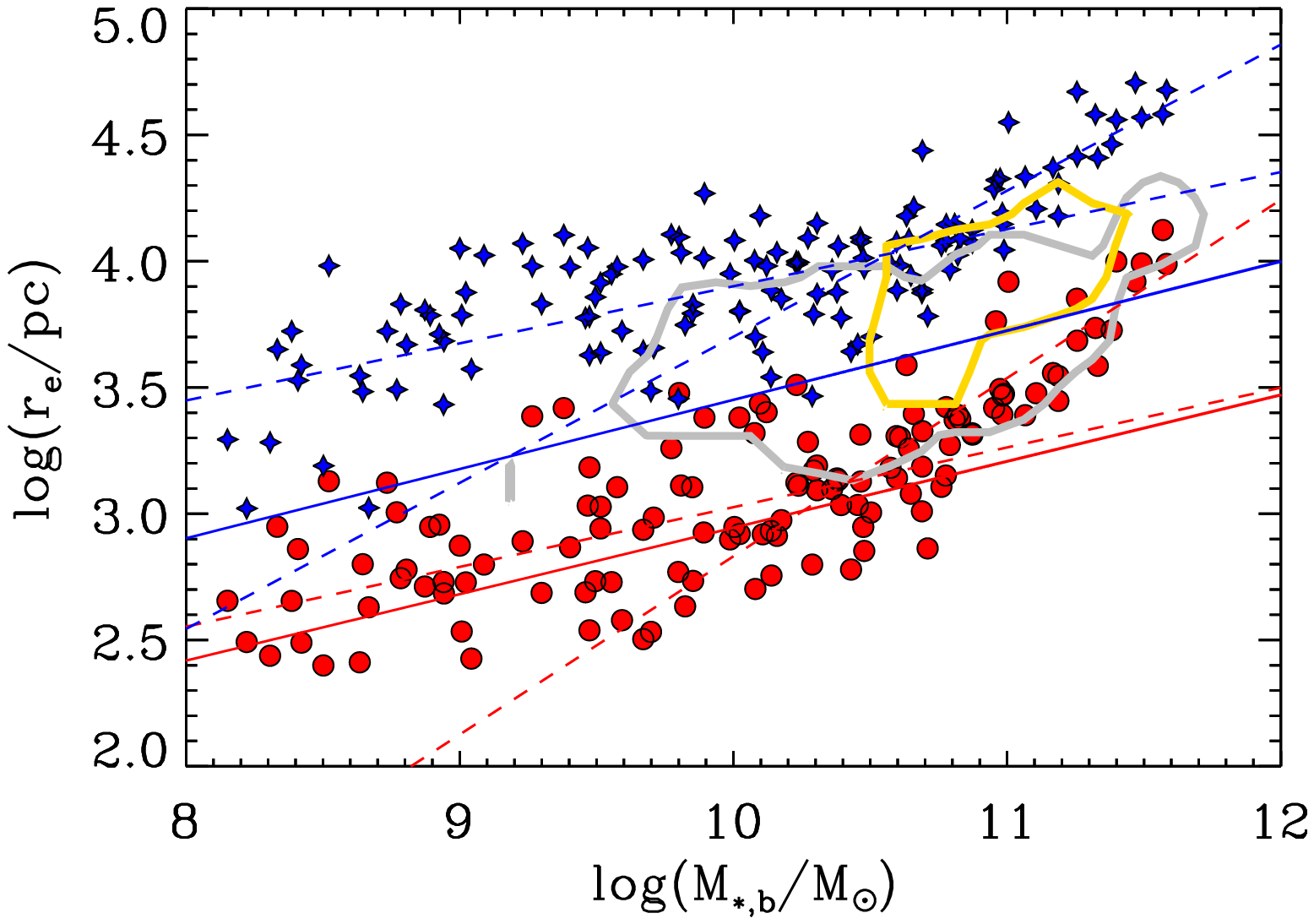}
    \caption{Mass-size relation for the bulges (circles) and discs (stars) of our sample galaxies. The stellar masses are those for the bulge component in both cases. The effective radii of both bulges and discs (1.678$\times h$) are obtained from the r-band photometric decompositions of \citet{mendezabreu17}. Black isocontours represents the mass-size relation of our galaxy sample without separating their structures. Gold isocontours show the mass-size relation of the photometrically defined {\it pure} ellipticals in the CALIFA sample \citep[see][for details]{mendezabreu18}. The blue and red solid lines show the best-fit obtained by \citet{lange16} for their sample of discs and spheroids, respectively. The blue and red dotted lines represent the best-fits obtained for our sample of discs and bulges, respectively. To perform our fit to the mass-size relation we divided the sample into two mass bins: $8 < \log{(M_{\star, \rm b}/M_{\sun})} < 10.5$ and $10.5 < \log{(M_{\star, \rm b}/M_{\sun})} < 12$ to capture the change in the slope.}
    \label{fig:mass-size-bulgemass}
\end{figure}

The mass-size relation shown in Fig.~\ref{fig:mass-size} provides key information about the galaxy-disc connection. The original idea about the formation of discs suggests that they are formed at the center of dark matter halos as a consequence of angular momentum conservation during the dissipational collapse of gas \citep{fallefstathiou80,mo98}. Therefore the stellar mass of the galaxy, which is linked to the mass of the halo, should be related to the size of the disc. This is exactly what is shown in the left panel of Fig.~\ref{fig:mass-size}, however this relation does not hold when the mass of the disc is considered alone (right panel), whereas it still does for bulges. The fact that it is not the mass of the discs what influences their morphology, but rather the mass of the galaxy, has important implications on their formation. First, it points towards a latter assembly of discs with respect to bulges, then favouring the inside-out view of galaxy evolution. In other words, if galaxy discs would have formed before their bulges, their stellar mass should correlate with their size, but this is not observed. A possibility might be that either angular momentum is not conserved during this early phase of disc formation or that the subsequent growth of a central bulge erases the correlation. A second implication is associated with the tight correlation of the disc size when considering the mass of the bulge component. This is shown in Fig.~\ref{fig:mass-size-bulgemass}. Again, our results point towards a scenario where the properties of bulges are tightly linked to the galaxy, and the size of discs is tightly linked to the bulge stellar mass. A scenario where bulges form out of galaxy mergers and discs are latter built from the remnant gas would fit into this picture. The merger would erase the original mass-size correlation of discs and the re-built disc size would mostly depend on the mass of the resultant bulge. A caveat to this interpretation is given in Fig.~\ref{fig:mass-size-evolution} since the mass-size relation for discs is not visible even at $z \sim 2$. 

%--------------------------- 
%---------------------------

The bulge influence on the disc properties does not seem to be reciprocal. Fig.~\ref{fig:n} shows  a clear correlation between the mass of the galaxy/bulge and their central concentration, expressed in terms of the S\'ersic index. However, there is no correlation between the mass of the disc and S\'ersic index. One could argue that the narrow range of available disc masses is playing against the correlation in this case, but we only show bins where the distribution of discs is statistically representative and there is no clear trend. Thus, our result indicates that the prominence of the disc is not related to the central concentration of the bulges. 

%------------------------------------------------------
\subsection{The morphological evolution of disc galaxies}

The picture emerging from our analysis of non-barred galaxies indicates that, most likely, bulges formed first with their properties being tightly bounded to those of the whole galaxy. Then galactic discs arise within a narrow mass range and with their properties being related to that of the bulge/galaxy, but not affecting them. A more detailed picture on the mass growth of both component will be explored in M\'endez-Abreu et al. (2020, in preparation). Here we will contextualise our results in terms of the evolution of their photometric properties.

A number of works have tried to establish the relative mass contribution of galactic bulges and discs at low redshift. Besides some differences that can be associated to the details of how structures are separated, the broad consensus is that stellar mass in galaxies is almost equally divided between spheroids and disc structures \citep{driver07,moffett16,thanjavur16}. Perhaps even more interesting is to know at which galaxy mass the systems are dominated by either the bulge or disc contribution. Fig.~\ref{fig:mass-fraction} shows a transition mass of $\log{(M_{\star, \rm b}/M_{\sun})} \sim 10.75$ for our sample. Therefore, one might be inclined to think that for galaxies over this transition mass, bulge related effects would drive galaxy evolution and disc related effect would do the same for lower masses. However, we have demonstrated in this paper that bulge properties, at all masses, are those most tightly related to the global galaxy whereas the disc component, even at low galaxy masses, do not correlate nor modify the properties of bulges.

Our results suggest that even low-mass bulges were likely formed at high redshift (see M\'endez-Abreu et al. 2020, in preparation, for a stellar population analysis) and their photometric properties have suffered little modifications during their evolution. This is in agreement with the scenario where massive high-redshift quiescent compact galaxies (red nuggets; \citealt{daddi05, damjanov09}) might end-up as bulges of of $z\sim0$ disc galaxies \citep{grahamscott13,delarosa16}. \citet{costantin20} suggested that this picture could be extended to the low-mass bulges of late-type galaxies, providing a general and broader picture of galaxy evolution at all masses in an inside-out scenario.  In a more recent paper, Costantin et al. (2021) analysed a sample of high-redshift bulge-to-disc galaxies showing that bulges formed in two waves; one at $z \sim6.2$ and another at $z \sim1.3$, suggesting again that bulges of disk-like galaxies harbor some of the oldest spheroids in the Universe. Our results are compatible with these ideas.  Another questions that arise are: which physical process/es lead/s to the central accumulation of gas necessary to build these low-mass compact galaxies at high redshift? Is this process different for high- and low-mass galaxies? At high masses, numerical simulations provide some answers. \citet{hopkins09} described how galaxy mergers are efficient mechanisms to fuel gas to the galaxy centers creating violent starburst. Then, depending on the gas fraction of the merging galaxies, their remnant could reform a disc. \citet{zolotov15} described the early formation of a gas-rich, star-forming disc fed by intense inflows and developing violent disc  instabilities. Then, the gaseous disc suffers a dissipative, quick compaction turning the galaxy into a compact, star-forming blue nugget that, immediately after the  compaction, suffers a fast quenching of star formation into a compact red nugget (see also \citet{birnboimdekel03}). Our observations suggest that galaxy discs observed at $z\sim0$ are formed after bulges. Nevertheless, they likely have little or no relation with the original discs formed before the compaction phase or with the merging event that ignited the formation of a bulge.

The morphological evolution of galaxies is generally interpreted in terms of the Hubble tuning-fork diagram, i.e., the scope is to understand the relative evolution of the bulge-to-disc mass contribution with respect to the global galaxy, which in turns defines the Hubble sequence. The facts that the disc evolution has a negligible effect on the properties of bulges (Fig.~\ref{fig:n}), the little evolution of the $B/T$ mass ratio with cosmic time (Fig.~\ref{fig:BT_evolution}, left panel), and the redshift evolution of the $B/D$ mass ratio (Fig.~\ref{fig:BT_evolution}, right panel), suggest that the Hubble type of a galaxy is mostly defined by the ability of the disc to grow around a pre-existing bulge, at least for those discs observed in the last 8 Gyrs. We have shown that for low-mass galaxies ($9.5 < \log{(M_{\star, \rm b}/M_{\sun})} < 10.5$ which mainly correspond to late type galaxies) the $B/D$ mass ratio does not change significantly with time, therefore the morphology (or Hubble type) of these galaxies is set-up at early stages of their evolution. Galaxies of larger masses undergo a phase of important disc growth that might lead to changes in their original morphology from earlier to later types. This suggested scenario might appear in conflict with the predicted scenarios of bulge growth from internal secular processes within galaxy discs \citep{kormendykennicutt04, athanassoula05}. Recent observational works have demonstrated the influence of bars on creating new central structures (Gadotti et al. 2020, submitted) and the high incidence of the so-called 'composite bulges' (a combination of bulges formed at high redshift and by secular evolution co-existing in the same galaxy) in barred systems \citep{mendezabreu14, erwin15}. Despite this, we cannot discard the presence of composite bulges in our analysis, the results presented here are representative of the 'main' bulge component in our galaxies. Regarding bar-related processes, acting in a secular evolution fashion, we remind here that our sample is bar free and therefore such processes should not have an effect on our sample. Notwithstanding this consideration, it is interesting to take into account the possibility that both branches of the Hubble tuning fork diagram would evolve differently. We will study the spectro-photometric properties of the structural components in barred galaxies in a future paper to investigate the prospects of this scenario.

%------------------------------------------------------
%------------------------------------------------------
\section{Conclusions}
\label{sec:conclusions}
In this first paper of the series, we describe the application of a new spectro-photometric decomposition code, called {\sc c2d}, to a sample of 129 non-barred galaxies from the CALIFA survey. The main advantage of our method is the ability to separate the bulge and disc contributions in the galaxy spectra in a spatially-resolved way. Therefore, the application of extensively tested algorithms to derive the stellar population properties from IFS is straightforward. To this aim, we use the {\sc Pipe3D} code which was initially conceived to analyse CALIFA data.

We show that our galaxy sample of non-barred galaxies is statistically representative of the nearby Universe in the magnitude range $-19.5 > M_r > -23$, which corresponds to galaxies in the stellar mass range $9.5 < \log{(M_{\star}/M_{\sun})} < 12$. We also compute the mass distribution function of bulges and discs finding that they are statistically representative of the nearby population in the  mass ranges $8 < \log{(M_{\star, \rm b}/M_{\sun})} < 11.5$ and $9.5 < \log{(M_{\star, \rm d}/M_{\sun})} < 11$, respectively.

We focus this paper in the analysis of the relations (and evolution) between the photometric properties of both bulges and discs and the stellar mass. We use the analysis of their stellar populations to find a strong relation between the $B/T$ mass ratio and the galaxy stellar mass. The evolution of the $B/T$ mass ratio with redshift (for a given galaxy mass) is very mild; however, we detect a clear increase of the disc component over the bulge in the cosmic evolution of the $B/D$ mass ratio. Therefore, this ratio seems more sensitive to the relative growth of both components with the most massive galaxies ($\log{(M_{\star}/M_{\sun})} > 10$) showing a larger disc increment since $z \sim 1$.

The mass-size relation contains crucial information about the formation process of galaxies, and we have been able to dissect this relation into their bulge and disc components. We find a clear upturn for both components at their most massive end ($\log{(M_{\star}/M_{\sun})} > 10.5$). The tight relation, and the upturn, holds for bulges when using their own stellar mass, but it disappears for galactic discs. This latter behaviour holds up to $z \sim 2$. In the case of bulges, this is caused by their negligible evolution in either size or mass. The evolution of discs is differential for high and low masses, using a disc stellar mass of $\log{(M_{\star, \rm d}/M_{\sun})} \sim 10.5$ for the separation. Low-mass discs show little evolution since $z \sim 2$ whereas the high-mass discs display a mass increase of a factor $\sim 2.3$ .

We also study the relation of the bulge S\'ersic index with the stellar mass of both the galaxy and individual components. Interestingly, we find a clear positive correlation between $n$ and the mass of both galaxy and bulge, but not with the mass of the discs.

All our results show a tight relation between the properties of bulges and their host galaxies. This connection has been largely debated and mainly described for massive galaxies; however, we found it is also valid for galaxies with stellar masses in the range $9.5 < \log{(M_{\star}/M_{\sun})} > 10.5$. Galactic discs in our sample show a different behaviour than bulges in their relation with global galaxy properties. First, they appear in a more narrow range of stellar masses, possibly indicating a mass-dependent efficiency in their formation, and second, the disc properties do not have an impact on the galaxy (or bulge) properties. The picture emerging from our results indicate that, most likely, bulges formed first with their properties being tightly bounded to those of the whole galaxy. Then, galactic discs arise with their properties being set up by those of the early bulge/galaxy, but not affecting them. We will further explore this scenario in the next papers of the series using a full analysis of the stellar mass growth and stellar populations gradients.

%------------------------------------------------------
%------------------------------------------------------
\section*{Acknowledgements}
JMA acknowledge support from the Spanish Ministerio de Economia y Competitividad (MINECO) by the grant AYA2017-83204-P. AdLC acknowledges support from grant AYA2016- 77237-C3-1-P from the Spanish Ministry of Economy and Competitiveness (MINECO). SFS thanks the projects ConaCyt CB-285080, FC-2016-01-1916 and PAPIIT IN100519. This paper is based on data from the Calar Alto Legacy Integral Field Area Survey, CALIFA, funded by the Spanish Ministery of Science under grant ICTS-2009-10, and the Centro Astron\'omico Hispano-Alem\'an. Based on observations collected at the Centro Astron\'omico Hispano Alem\'an (CAHA) at Calar Alto, operated jointly by the  Max-Planck Institut f\"ur Astronomie and the Instituto de Astrof\'isica de Andaluc\'ia.

\section*{Data availability}
The data underlying this article are available in the article and in its online supplementary material.



%\bibliographystyle{mnras}

\bibliographystyle{mnras}
\bibliography{reference}

%%%%%%%%%%%%%%%%%%%%%%%%%%%%%%%%%%%%%%%%%%%%%%%%%%

%%%%%%%%%%%%%%%%% APPENDICES %%%%%%%%%%%%%%%%%%%%%

%\appendix

%%%%%%%%%%%%%%%%%%%%%%%%%%%%%%%%%%%%%%%%%%%%%%%%%%

% Don't change these lines
\bsp	% typesetting comment
\label{lastpage}
\end{document}